\documentclass[%
reprint,
superscriptaddress,
amsmath,amssymb,
aps,
]{revtex4-1}

\usepackage{booktabs}
\usepackage{multirow}
\usepackage{subcaption}
\usepackage{amsmath}
\usepackage{graphics}
\usepackage{mathtools}
\usepackage{booktabs}
\usepackage{siunitx}

\usepackage{bm} 
\usepackage{graphicx}
\usepackage{dcolumn}
\usepackage{bm}
\usepackage{hyperref}

\begin{document}

\preprint{APS/123-QED}

\title{Distinguishing Between Dark Matter-Black Hole Systems and Naked Singularities via Quasi-Periodic Oscillations   }

\author{Zheng Ma}

\author{Zhaoyi Xu}

\author{Meirong Tang}%
\email{tangmr@gzu.edu.cn(Corresponding author)}
\affiliation{%
 College of Physics,Guizhou University,Guiyang,550025,China
}%


\begin{abstract}
Quasi-Periodic Oscillations (QPOs) are an important phenomenon commonly observed in the X-ray radiation of black holes and neutron stars, closely related to the dynamics of accretion disks around compact objects and general relativistic effects. The objective of this study is to use the QPO phenomenon to distinguish between dark matter-black hole systems and naked singularities, as well as to investigate the effects of different dark matter models (Cold Dark Matter, CDM, and Scalar Field Dark Matter, SFDM) on the accretion disk dynamics.By introducing a dark matter correction model within the framework of general relativity, we systematically investigate the differences in dragging effects, characteristic frequency distribution, and the innermost stable circular orbit (ISCO) radius between dark matter-black hole systems and naked singularities, while analyzing the potential coupling between QPO frequencies and dark matter distribution.The main results of this study are as follows: $\nu_r$ and $\nu_\theta$ in dark matter-black hole systems can be identified as HFQPOs, while for lower spins ($a < 0.5$), $\nu_\text{nod}$ can be identified as LFQPOs, and for higher spins ($1 > a \geq 0.5$), $\nu_\text{nod}$ falls within the HFQPO observation range. Cold Dark Matter (CDM) and Scalar Field Dark Matter (SFDM) modulate the accretion disk dynamics at the order of $10^{-6}$.

\begin{description}
\item[Keywords]
Scalar Field Dark Matter, Cold Dark Matter, Quasi-Periodic Oscillations, innermost stable circular orbit (ISCO), Kerr Helo.
\end{description}
\end{abstract}

\maketitle


\section{\label{sec:level1}Introduction}
Black holes are among the most extreme spacetime regions in the universe and represent crucial predictions of general relativity. The Schwarzschild solution first quantitatively described their spacetime structure in spherical symmetry, while the Kerr solution revealed the complex spacetime geometry of rotating black holes\cite{kerr1963gravitational}. The 2019 direct imaging of the supermassive black hole in M87's galactic center by the Event Horizon Telescope\cite{event2019first,akiyama2022first} marked the observational confirmation of black holes, transitioning them from theoretical entities to empirical realities. However, another profound implication of general relativity—the singularity problem—remains unresolved. To address the physical paradoxes caused by singularities, Penrose proposed the cosmic censorship conjecture, which posits that singularities should always be concealed by event horizons to preserve causal structure. Nevertheless, numerical simulations suggest potential formations of naked singularities under extreme angular momentum conditions, non-spherical collapse scenarios, or quantum gravitational effects e.g.\cite{JoshiMalafarina2011,Ghosh2024,ShapiroTeukolsky1991,Hawking1974,EmparanFabbriKaloper2002}. The exposure of such singularities could lead to the breakdown of spacetime causality and challenge the applicability of established physical laws. Therefore, distinguishing between black holes and naked singularities constitutes not only a foundational problem in gravitational theory but also a critical pathway toward exploring quantum gravity.

On cosmological scales, dark matter constitutes 26.8\% \cite{Planck2020}of the universe's mass-energy content, with its gravitational effects dominating galaxy formation and evolution. Although dark matter has not been directly observed, substantial evidence supports its existence through galaxy rotation curves, such as galactic dynamics, and cosmic microwave background radiation and so on e.g.\cite{Planck2020,Rubin1980,NFW1996}. The concept was initially proposed to explain observational anomalies including galactic rotation profiles of disk galaxies, velocity dispersions in galaxy clusters, and mass-to-light ratios. Subsequent studies revealed dark matter's crucial role in early universe structure formation and gravitational lensing phenomena. Among various dark matter models, cold dark matter (CDM) remains the most prominent, typically postulated as a collisionless particle system in cosmological standard models. While successful in explaining large-scale structure formation e.g.\cite{Springel2005}, CDM faces challenges in resolving the core-cusp problem in dwarf galaxy rotation curves e.g.\cite{deBlok2010} and the missing satellite problem \cite{Klypin1999}. Alternative models including self-interacting dark matter (SIDM) e.g.\cite{Spergel2000,Schive2014, Rocha2013}, fuzzy dark matter e.g.\cite{Du2017}, and scalar field dark matter (SFDM) e.g.\cite{Matos2001,Robles2013} and so on have been proposed. The SFDM model, incorporating self-coupled scalar fields, effectively alleviates observational tensions at galactic scales. Notably, dynamical coupling between dark matter halos and compact objects may significantly influence accretion disk structures e.g.\cite{Cardoso2017,Berezhiani2018}, offering unique opportunities to probe dark matter properties through extreme astrophysical environments.
For compact objects interacting with dark matter, quasi-periodic oscillations (QPOs) exhibit model-dependent behaviors: Herrera-Aguilar and Malafarina demonstrated enhanced low-frequency QPOs under SFDM models\cite{HerreraAguilar2023}, while Dihingia et al. reported high-frequency QPO suppression in SIDM scenarios\cite{Dihingia2021}. Distinct QPO signatures also emerge around different singularity types – Kološ et al. revealed characteristic twin-peak QPO spectra from accretion flows near naked singularities through GRMHD simulations\cite{Kolos2020}, and Jusufi et al. identified significantly higher Lense-Thirring precession frequencies 
 around naked singularities compared to black holes of equal mass\cite{Jusufi2022}. This work systematically investigates QPO frequency dependencies on black hole/naked singularity nature and dark matter properties, employing spacetime models modified by CDM and SFDM from \cite{xu2018black}. Specific investigations include: (1) Differences in accretion disk precession behaviors around black holes and naked singularities under CDM/SFDM frameworks; (2) Sensitivity of QPO characteristic frequencies to dark matter parameters (scalar field mass, self-coupling strength).

The paper is structured as follows: Section II examines critical spin parameter \(a\), event horizons, and ergoregions for Kerr black holes. Section III derives precession frequency formulas including Lense-Thirring frequencies (outside ergoregions) and generalized precession expressions, with graphical analysis of gyroscopic precession behaviors. Section IV discusses innermost stable circular orbits (ISCO) and their characteristic frequencies. Section V establishes connections between ISCO frequencies and observed QPOs. Section VI presents conclusions.Throughout this paper, we adopt the geometric unit system with \( c = G = M = 1 \). The metric signature used is \((- , +, +, +)\).

\section{\label{sec:level2}Black hole within the dark matter halo}
\subsection{spacetime metric}
The existence of dark matter exerts significant influence on the investigation of spacetime characteristics near black holes/naked singularities. In this paper, we adopt the Kerr-like spacetime metrics coupled with cold dark matter (CDM) and scalar field dark matter (SFDM) from \cite{xu2018black}, where the corresponding spacetime line elements are respectively expressed as:

CDM case:
\begin{align}
	ds^2 =& -\left(1 - \frac{r^2 + \frac{2GM}{c^2}r - r^2 \left[ 1 + \frac{r}{R_s} \right]^{-\frac{8\pi G \rho_c R_s^3}{c^2 r}}}{\Sigma^2} \right) dt^2 
	\nonumber \\
	& + \frac{\Sigma^2}{\Delta} dr^2 + \Sigma^2 d\theta^2 
	\nonumber \\
	& + \frac{\sin^2\theta}{\Sigma^2} \left( \left(r^2 + a^2\right)^2 - a^2 \Delta \sin^2\theta \right) d\phi^2
	\nonumber \\
	& - \frac{2 a \sin^2\theta}{\Sigma^2} \left(r^2 + \frac{2GM}{c^2}r \right. 
	\nonumber \\
	& \left. - r^2 \left[1 + \frac{r}{R_s} \right]^{-\frac{8\pi G \rho_c R_s^3}{c^2 r}}\right) d\phi \, dt,
	\label{eq:1}
\end{align}
where
\begin{align}
	\Delta = r^2 \left[1 + \frac{r}{R_s}\right]^{-\frac{8\pi G \rho_c R_s^3}{c^2 r}} - \frac{2GM}{c^2}r + a^2.
	\label{eq:2}
\end{align}

SFDM case:
\begin{align}
	ds^2 = & -\left( 1 - \frac{r^2 + \frac{2 G M}{c^2} r - r^2 \exp\left[ -\frac{8 G \rho_c R_s^2}{\pi} \frac{\sin\left(\frac{\pi r}{R_s}\right)}{\frac{\pi r}{R_s}} \right]}{\Sigma^2} \right) dt^2 \notag \\ 
	& + \frac{\Sigma^2}{\Delta} dr^2 + \Sigma^2 d\theta^2 \notag \\
	& + \frac{\sin^2\theta}{\Sigma^2} \left( (r^2 + a^2)^2 - a^2 \Delta \sin^2\theta \right) d\phi^2 \notag \\
	& - \frac{2 a \sin^2\theta}{\Sigma^2} \left( r^2 + \frac{2 G M}{c^2} r \right. \notag \\
	& \left. - r^2 \exp\left[ -\frac{8 G \rho_c R_s^2}{\pi} \frac{\sin\left(\frac{\pi r}{R_s}\right)}{\frac{\pi r}{R_s}} \right] \right) d\phi \, dt ,\label{eq:3}
\end{align}
where
\begin{align}
	\Delta = r^2 \exp\left[ -\frac{8G\rho_c R_s^2}{\pi} \frac{\sin\left(\frac{\pi r}{R_s}\right)}{\frac{\pi r}{R_s}} \right] - \frac{2GM}{c^2}r + a^2,
	\label{eq:4}
\end{align}
where \( R_s \) is the scale radius of the dark matter halo, \( \rho_c \) refers to the central density of the dark matter halo, \( G \) is the gravitational constant, \( c \) is the speed of light. In the subsequent discussion, we adopt the natural unit system (\( G = c = 1 \)) to simplify the spacetime metrics under two dark matter models, and unless otherwise specified, we assume by default that the mass \( M = 1 \), i.e., in units of the black hole mass. The scale radius and central density of the dark matter halo come from the low surface brightness galaxy ESO1200211 [17,18]. For CDM: \( \rho_c = 2.45 \times 10^{-3}\, M_\odot/\text{pc}^3 \), \( R_s = 5.7\, \text{kpc} \), for SFDM: \( \rho_c = 13.66 \times 10^{-3}\, M_\odot/\text{pc}^3 \), \( R_s = 2.92\, \text{kpc} \), \( a \) is the black hole spin parameter. According to the physical definition of the black hole horizon, these can be given by the following structural equations,respectively:

CDM case:
\begin{align}
\Delta = r^2 \left[1 + \frac{r}{R_s}\right]^{-(8\pi G \rho_c R_s^3)/(c^2 r)} - \frac{2 G M r}{c^2} + a^2 = 0.
\label{eq:5}
\end{align}

SFDM case:
\begin{align}
	\Delta = r^2 \exp\left[-\frac{8 G \rho_c R_s^2}{\pi} \frac{\sin\left(\frac{\pi r}{R_s}\right)}{\frac{\pi r}{R_s}}\right] - \frac{2 G M r}{c^2} + a^2 = 0.
	\label{eq:6}
\end{align}
\subsection{The event horizon of a black hole}
A black hole possesses an event horizon boundary. When the black hole horizon disappears, the naked singularity at the center becomes exposed. From Eq.\ref{eq:5} and Eq.\ref{eq:6}, we observe that the position of the event horizon depends on the black hole spin parameter \(a\). Through equation transformation, we obtain the relationship between \(a^2\) and \(r\) as follows:

CDM Case:
\begin{align}
	a^2 = & -r^2 \left[1 + \frac{r}{R_s}\right]^{-\frac{8\pi\rho_c R_s^3}{r}} + 2r \notag \\
	\frac{da^2}{dr} = & -r^2 \exp\left[-\frac{8\rho_c R_s^2}{\pi} \frac{\sin\left(\frac{\pi r}{R_s}\right)}{\frac{\pi r}{R_s}}\right] \notag \\
	& \times \left( \frac{8\rho_c R_s^3 \sin\left(\frac{\pi r}{R_s}\right)}{\pi^2 r^2} - \frac{8\rho_c R_s^2 \cos\left(\frac{\pi r}{R_s}\right)}{\pi r} \right) + 2 \label{eq:7} 
\end{align}

SFDM case:
\begin{align}
	a^2 = & -r^2 \exp\left[-\frac{8\rho_c R_s^2}{\pi} \frac{\sin\left(\frac{\pi r}{R_s}\right)}{\frac{\pi r}{R_s}}\right] + 2r \notag \\
	\frac{da^2}{dr} = & -r^2 \exp\left[-\frac{8\rho_c R_s^2}{\pi} \frac{\sin\left(\frac{\pi r}{R_s}\right)}{\frac{\pi r}{R_s}}\right] \notag \\
	& \times \left( \frac{8\rho_c R_s^3 \sin\left(\frac{\pi r}{R_s}\right)}{\pi^2 r^2} - \frac{8\rho_c R_s^2 \cos\left(\frac{\pi r}{R_s}\right)}{\pi r} \right) + 2 \label{eq:8}
\end{align}
We analyze three scenarios of the horizon function based on the maximum value \(a_{\text{max}}^2\) of \(a^2\). 
\begin{itemize}
	\item When \(a^2 < a_{\text{max}}^2\), the horizon function has two distinct solutions, indicating that the black hole possesses two horizons.
	\item When \(a^2 = a_{\text{max}}^2\), the horizon function has a single solution, corresponding to an extremal black hole in the metric.
	\item When \(a^2 > a_{\text{max}}^2\), the horizon function has no solution, implying that the metric describes a naked singularity.
\end{itemize}

\begin{figure}[ht]
	\centering
	\includegraphics[width=0.5\textwidth]{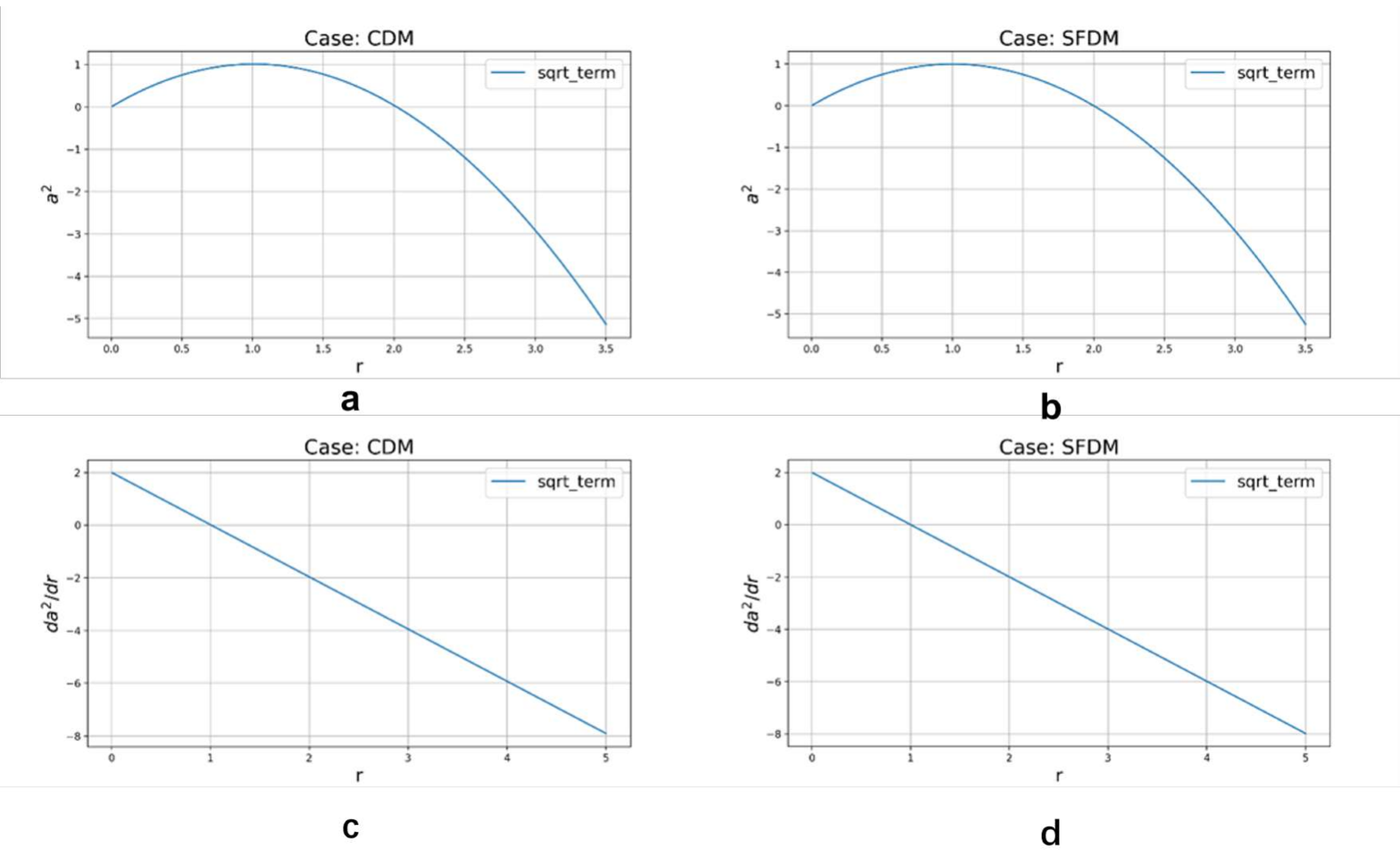}
	\caption{The variations of \( a^2 \) and \(\frac{d(a^2)}{dr}\) as functions of \( r \) in both Cold Dark Matter (CDM) and Scalar Field Dark Matter (SFDM) scenarios are illustrated.
	}
	\label{fig:1}
\end{figure}

As shown in Fig.\,1, the maximum value of $a$ is 1 (specifically, when $a > 1$). For $a > 1$, the spacetime metrics in Eq.\ref{eq:1} and Eq.\ref{eq:3} represent a naked singularity; for $a < 1$, they describe a black hole; while $a = 1$ corresponds to an extremal black hole.

\subsection{Ergosphere}
A rotating black hole is surrounded by a unique region called the Ergosphere, bounded by the event horizon and the static limit surface. Within this region, spacetime is dragged by the black hole's spin, resulting in the frame-dragging effect (Lense-Thirring effect). All matter (including photons) cannot remain stationary here, and it is also the domain where the Penrose process occurs. According to \cite{chakraborty2014strong,straumann2013general}, the Lense-Thirring (LT) precession frequency in Kerr spacetime can be derived within this ergosphere. The outer boundary of this region is defined by the static limit surface, given by:
\begin{align}
	g_{tt}=0.\label{eq:9}
\end{align}
decide,for CDM case:
\begin{align}
	-\left( 1 - \frac{r^2 + \frac{2GM r}{c^2} - r^2 \left[ 1 + \frac{r}{R_s} \right]^{-\frac{8\pi G \rho_c R_s^3}{c^2 r}}}{\Sigma^2} \right) = 0.\label{eq:10}
\end{align}

SFDM case:
\begin{align}
	-\left( 1 - \frac{r^2 + \frac{2GM r}{c^2} - r^2 \exp\left[-\frac{8 G \rho_c R_s^2}{\pi} \frac{\sin\left(\frac{\pi r}{R_s}\right)}{\frac{\pi r}{R_s}}\right]}{\Sigma^2} \right) = 0.\label{eq:11}
\end{align}
Due to the complexity of Eq.\ref{eq:10} and Eq.\ref{eq:11}, it is difficult to obtain an analytical expression for the static limit surface. Therefore, we employ numerical methods to determine the position of the static limit surface. Of course, in the case of a naked singularity, the definition of the ergosphere and static limit surface no longer applies, but we still plot the behavior of the metric component $g_{tt}$ when the singularity is a naked singularity, and its significance will be discussed in Section 3.
\begin{figure}[htbp]
	\centering
	\begin{minipage}[b]{0.5\textwidth}
		\includegraphics[width=\textwidth, height=5cm, keepaspectratio]{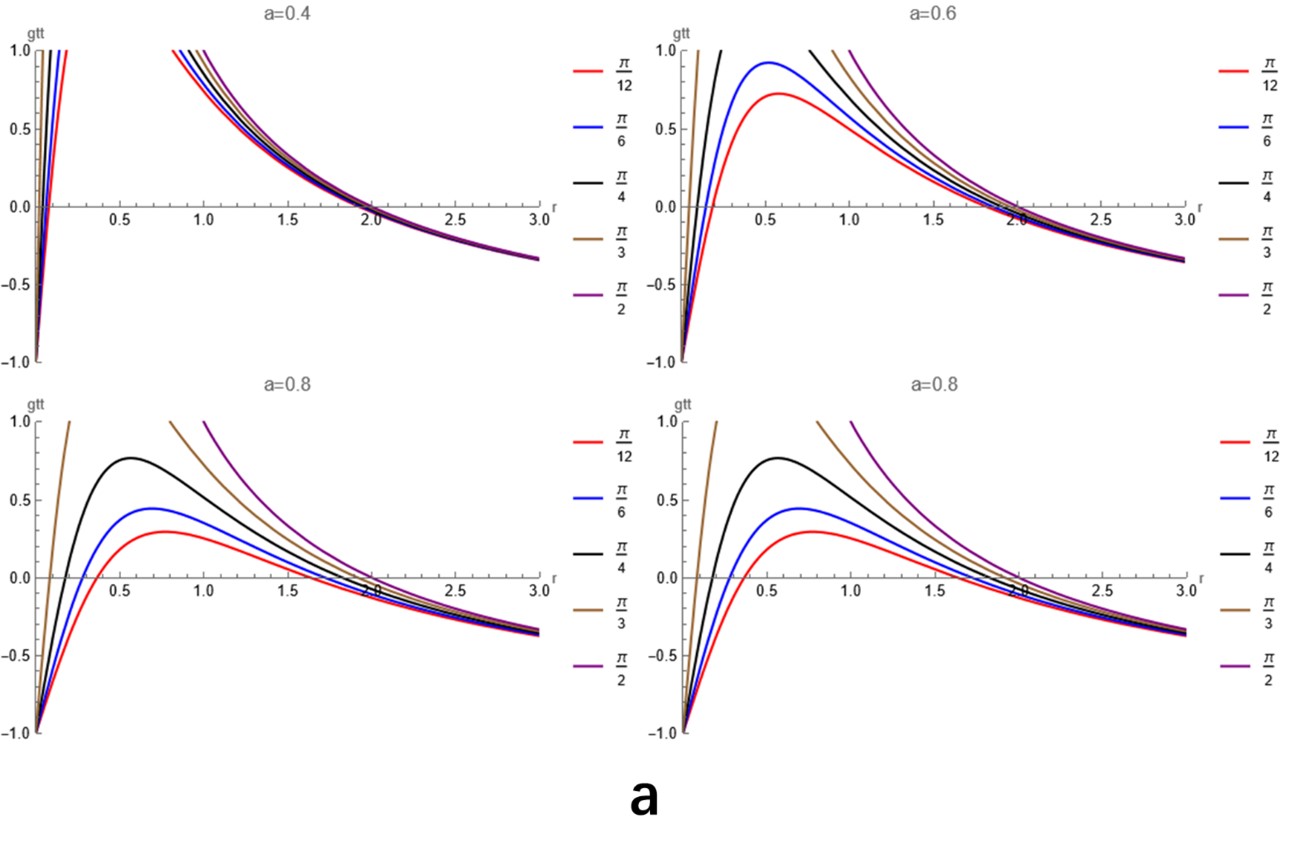}
	\end{minipage}
	\hfill
	\begin{minipage}[b]{0.5\textwidth}
		\includegraphics[width=\textwidth, height=5cm, keepaspectratio]{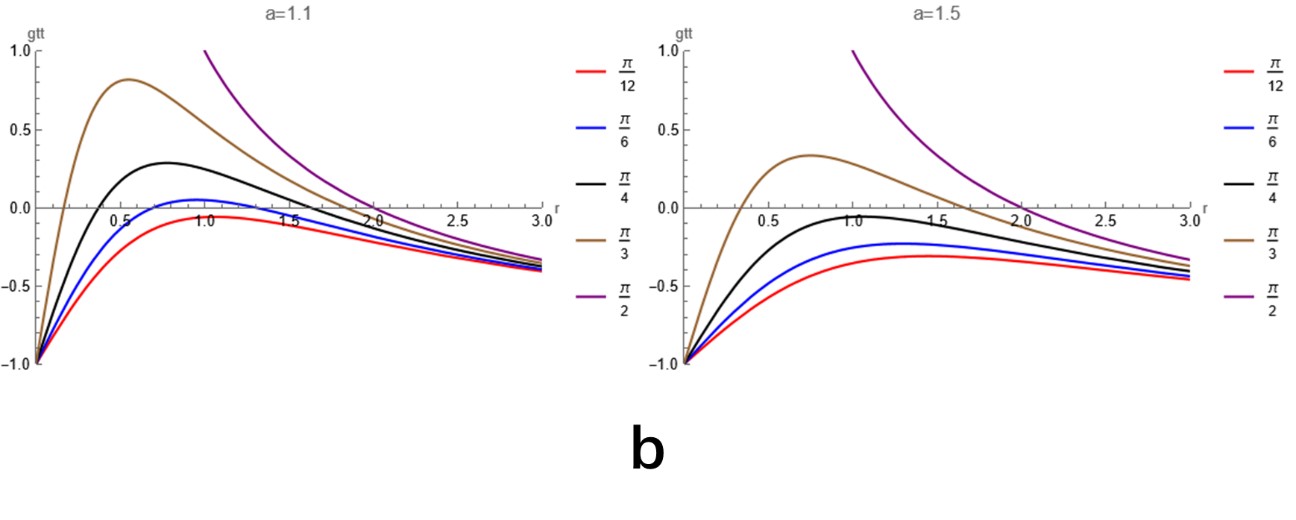}
	\end{minipage}
	\hfill
	\begin{minipage}[b]{0.5\textwidth}
		\includegraphics[width=\textwidth, height=5cm, keepaspectratio]{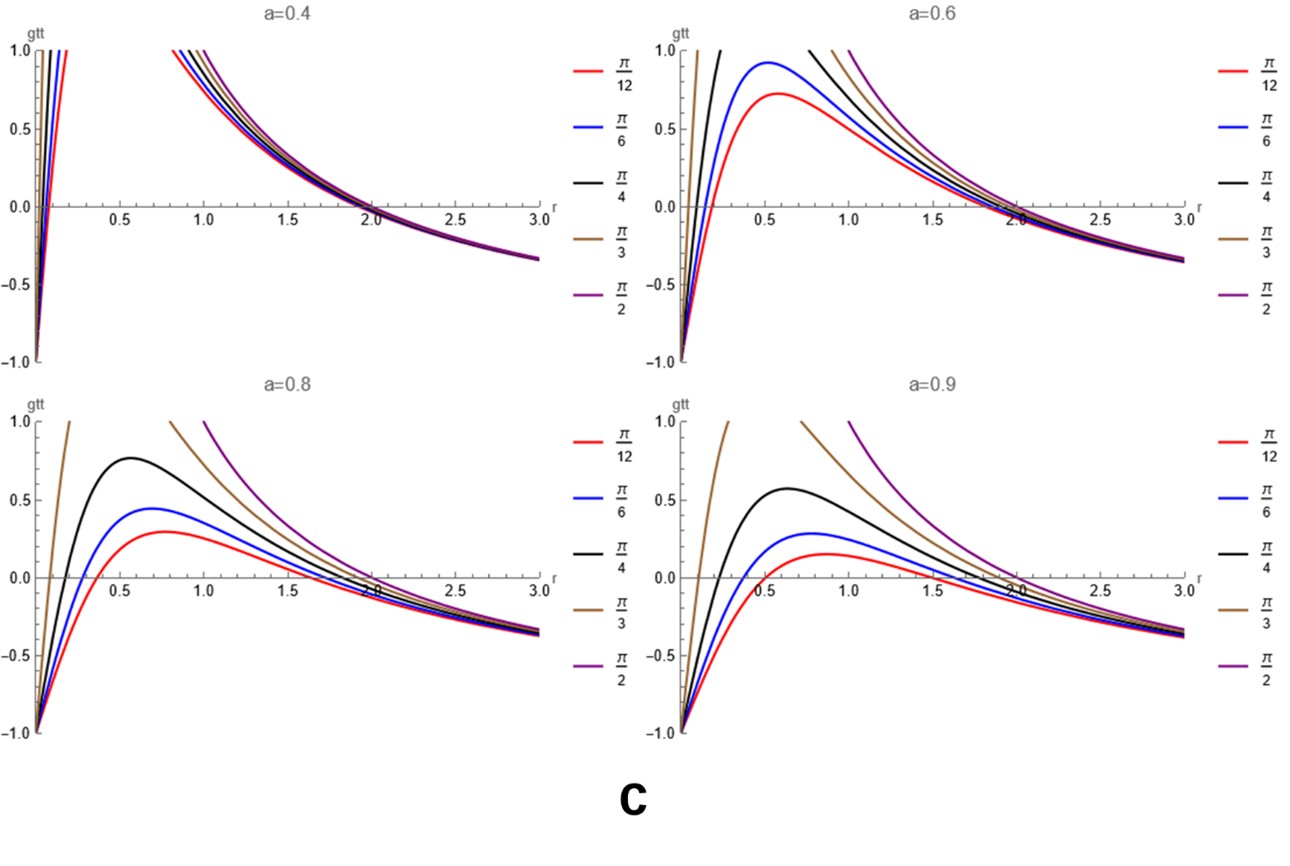}
	\end{minipage}
	\hfill
	\begin{minipage}[b]{0.5\textwidth}
		\includegraphics[width=\textwidth, height=5cm, keepaspectratio]{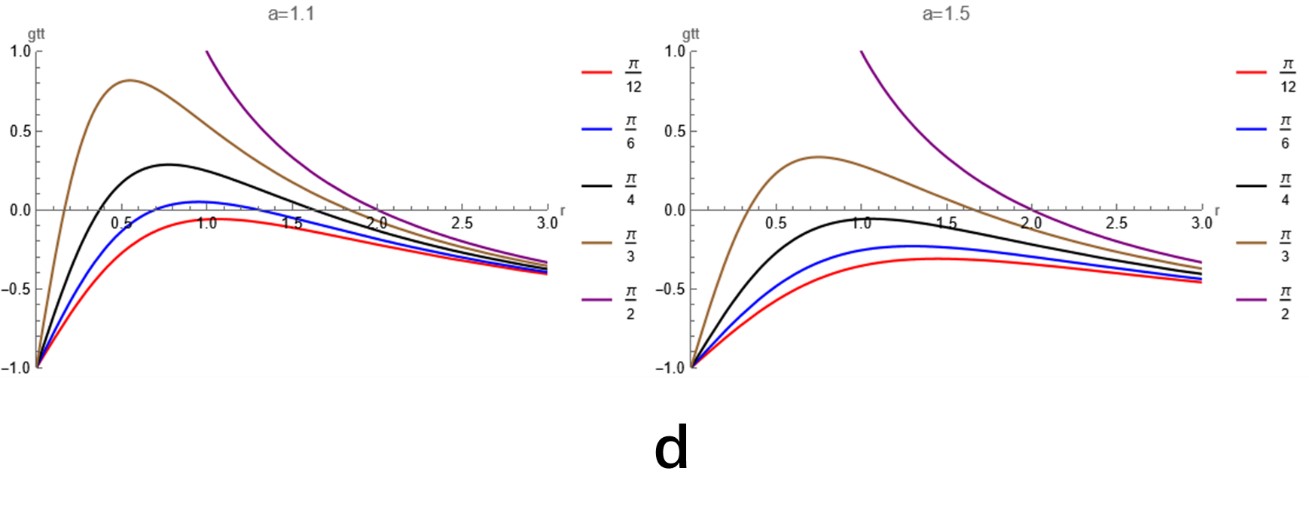}
	\end{minipage}
	\caption{The values of the metric component \( g_{tt} \) are plotted for both CDM (a, b) and SFDM (c, d) scenarios, where the intersections of the curves with the vertical coordinate axis indicate the positions of the static limit surfaces.}.\label{fig:2}
\end{figure}

\section{\label{sec:level3}Precession Frequency}
In this section, we examine the frame-dragging effect in Kerr-like black holes under two scenarios---Cold Dark Matter (CDM) and Scalar Field Dark Matter (SFDM)---by studying the spin behavior of test gyroscopes carried by static observers. The computational procedure adheres to the standard method outlined by\cite{Chandrasekhar1983,straumann2013general}. If a static observer is situated in a spacetime possessing a timelike Killing vector field ($K$), and this observer moves along the integral curve $\gamma(\tau)$ of $K$, the corresponding four-velocity $u^\mu$ is expressed as:

\begin{align}
u^\mu = \mu_{\text{obs}}^{\alpha} = \mu_{\text{obs}}^{t} = (1, 0, 0, 0).\label{eq:12}
\end{align}
and
\begin{equation}
	\mu = \left( -\langle K, K \rangle \right)^{-1/2} K.\label{eq:13}
\end{equation}
In any stationary spacetime, we may choose the Killing vector \( K \) in the form \( K = \partial / \partial t \). If a set of basis vectors \( \{e_i\} \) orthogonal to the curve \( \gamma \) is selected, they satisfy the condition of Lie transport along \( K \):

\begin{align}
	L_K e_i &= 0.\label{eq:14}
\end{align}

This parallel transport implies that \( e_i \) remains constant along the motion of \( \gamma \). Since the observer is at rest relative to the frame, the reference frame constituted by \( \{e_i\} \) is a stationary reference frame for an observer moving along \( \gamma \). For arbitrary vectors \( X \) and \( Y \), according to the parallel transport condition,

\begin{align}
	L_K g(X, Y) &= K \langle X, Y \rangle - \langle L_K X, Y \rangle - \langle X, L_K Y \rangle = 0.\label{eq:15}
\end{align}

When the condition \( L_K X = L_K Y = 0 \) holds, the orthogonality of vectors \( X \) and \( Y \) remains preserved. Thus, for \( i = 1, 2, 3 \), \( e_i \) is perpendicular to the plane formed by \( K \) and \( \mu \). For any two vectors \( e_i \) and \( e_j \), their rotation is computed via the differential:

\begin{align}
	\omega_{ij} &= (-\langle K, K \rangle)^{-1/2} \langle e_j, \nabla_{e_i} K \rangle.\label{eq:16}
\end{align}

Assuming the connection is torsion-free (the black hole spacetime considered in this paper satisfies the torsion-free condition), i.e.,

\begin{align}
	T(K, e_i) &= \nabla_K e_i - \nabla_{e_i} K - [K, e_i] = 0.\label{eq:17}
\end{align}

it follows that \( \nabla_K e_i = \nabla_{e_i} K \). Utilizing the antisymmetry property of the Killing vector \( K \), Eq.\ref{eq:16} can be simplified to:

\begin{align}
	\omega_{ij} &= \left[ \frac{1}{2} (-\langle K, K \rangle) \right]^{-1/2} \left( \nabla_{e_i} K (e_j) - \nabla_{e_j} K (e_i) \right).\label{eq:18}
\end{align}

Alternatively, this can be expressed as:

\begin{align}
	\omega_{ij} &= \left[ \frac{1}{2} (-\langle K, K \rangle) \right]^{-1/2} dK (e_i, e_j).\label{eq:19}
\end{align}

Using the Hodge star operator,Eq.\ref{eq:18} can be generalized to a global rotation measure in spacetime, written as:

\begin{align}
	\Omega_p &= \left[ \frac{1}{2} (-\langle K, K \rangle) \right]^{-1} *(K \wedge dK).\label{eq:20}
\end{align}

In a general static spacetime, we may choose \( K = \partial_0 \), and in this case, Eq.\ref{eq:20} represents the Lense-Thirring (LT) precession frequency. If \( K^2 = g_{00} \), then:

\begin{align}
	dK &= g_{00,k} \, dx^k \wedge dx^0 + g_{0i,k} \, dx^k \wedge dx^i.\label{eq:21}
\end{align}

Substituting into the Hodge star operator, we obtain:

\begin{align}
	*(K \wedge dK) &= g_{00}^2 \left( \frac{g_{0i}}{g_{00}} \right)_{,j} *(dt \wedge dx^j \wedge dx^i) \notag \\
	&\quad + g_{0k} g_{0i,j} *(dx^k \wedge dx^j \wedge dx^i)\label{eq:22}
\end{align}
Using the relation \( *(dt \wedge dx^j \wedge dx^i) = -\frac{1}{\sqrt{-g}} \epsilon_{ijl} g_{lu} dx^u \), the precession frequency can be simplified to:

\begin{align}
	\Omega_{\text{LT}} &= \frac{g_{00}}{2 \sqrt{-g}} \epsilon_{ijl} \left( \frac{g_{0i}}{g_{00}} \right)_{,j} \left( g_{lk} - \frac{g_{l0} g_{k0}}{g_{00}} \right) dx^k,\label{eq:23}
\end{align}

which can be further expressed as:

\begin{align}
	\Omega_{\text{LT}} &= \frac{g_{00}}{2 \sqrt{-g}} \epsilon_{ijl} \left( \frac{g_{0i}}{g_{00}} \right)_{,j} \left( \partial_l - \frac{g_{l0}}{g_{00}} \partial_0 \right).\label{eq:24}
\end{align}

In an axisymmetric stationary spacetime, when the observer is located on the ergosphere surface, it follows that, since \( g_{00} = 0 \), \( \Omega \) becomes divergent. However, once the observer enters the ergoregion, the Killing vector \( K \) is no longer timelike but becomes spacelike due to the frame-dragging effect caused by the black hole's rotation, preventing the observer from remaining stationary. In this state, the frame-dragging effect imparts an angular velocity \( \Omega \) to the observer, although \( r \) and \( \theta \) remain constant in polar coordinates. Consequently, the four-velocity is:

\begin{align}
	\mu = \mu_{\text{obs}}^{\alpha} &= \mu_{\text{obs}}^{t} = (1, 0, 0, \Omega).\label{eq:25}
\end{align}

Within and on the surface of the ergoregion, the Killing vector can be decomposed into a linear combination of a timelike part \( \partial_0 \) and a spacelike part \( \partial_c \):

\begin{align}
	K &= \partial_0 + \Omega \partial_c,\label{eq:26}
\end{align}

indicating that this spacetime is independent of \( x^0 \) and \( x^c \). The covariant form of \( K \) can be written as:

\begin{align}
	\tilde{K} &= g_{0\nu} dx^\nu + \Omega g_{\gamma c} dx^\gamma,\label{eq:27}
\end{align}

where \( \gamma, \nu = 0, 1, 2, 3 \). Substituting into the spacetime metric and separating components:

\begin{align}
	\tilde{K} &= (g_{00} dx^0 + g_{0c} dx^c + g_{0i} dx^i) + \Omega (g_{0c} dx^0 + g_{cc} dx^c + g_{ic} dx^i),\label{eq:28}
\end{align}

where \( i = 2, 3 \). Inside the ergoregion, focusing on components relevant to frame-dragging (\( g_{00}, g_{0c}, g_{cc} \)) and neglecting \( g_{0i}, g_{ic} \), we obtain:

\begin{align}
	\tilde{K} &= (g_{00} dx^0 + g_{0c} dx^c) + \Omega (g_{0c} dx^0 + g_{cc} dx^c).\label{eq:29}
\end{align}

The corresponding \( d\tilde{K} \) is:

\begin{align}
	d\tilde{K} &= g_{00,k} dx^k \wedge dx^0 + g_{0c,k} dx^k \wedge dx^c\notag\\
	& + \Omega (g_{0c,k} dx^k \wedge dx^0 + g_{cc,k} dx^k \wedge dx^c).\label{eq:30}
\end{align}
According to Eq.~\ref{eq:20}, substituting Eq.~\ref{eq:29} and Eq.~\ref{eq:30}, the expression for the precession frequency is given by:
\begin{align}
	(\tilde{\Omega}_p) = \frac{\varepsilon_{ckl} g_{lu} dx^u}{2\sqrt{-g} \left(1 + 2\Omega \frac{g_{0c}}{g_{00}} + \Omega^2 \frac{g_{cc}}{g_{00}}\right)} \notag \\
	\times \left[ \left( g_{0c,k} - \frac{g_{0c}}{g_{00}} g_{00,k} \right) + \Omega \left( g_{cc,k} - \frac{g_{cc}}{g_{00}} g_{00,k} \right) \right. \notag \\
	\left. + \Omega^2 \left( \frac{g_{0c}}{g_{00}} g_{cc,k} - \frac{g_{cc}}{g_{00}} g_{0c,k} \right) \right].\label{eq:31}
\end{align}
Here, we adopt \( K^2 = g_{00} + 2\Omega g_{0c} + \Omega^2 g_{cc} \). In a stationary axisymmetric spacetime with polar coordinates \( (0, r, \theta, \phi) \), the expression for \( (\tilde{\Omega}_p) \) can be simplified as:
\begin{align}
	\Omega_p = \frac{\varepsilon_{ckl}}{2\sqrt{-g} \left(1 + 2\Omega \frac{g_{0c}}{g_{00}} + \Omega^2 \frac{g_{cc}}{g_{00}}\right)} \notag \\
	\times \Bigg[ \sqrt{g_{\theta\theta}} \left[ \left( g_{0\phi,r} - \frac{g_{0\phi}}{g_{00}} g_{00,r} \right) \right. \notag \\
	+ \Omega \left( g_{\phi\phi,r} - \frac{g_{\phi\phi}}{g_{00}} g_{00,r} \right) \notag \\
	+ \Omega^2 \left( \frac{g_{0\phi}}{g_{00}} g_{\phi\phi,r} - \frac{g_{\phi\phi}}{g_{00}} g_{0\phi,r} \right) \Big] \partial_\theta \notag \\
	- \sqrt{g_{rr}} \left[ \left( g_{0\phi,\theta} - \frac{g_{0\phi}}{g_{00}} g_{00,\theta} \right) \right. \notag \\
	+ \Omega \left( g_{\phi\phi,\theta} - \frac{g_{\phi\phi}}{g_{00}} g_{00,\theta} \right) \notag \\
	+ \Omega^2 \left( \frac{g_{0\phi}}{g_{00}} g_{\phi\phi,\theta} - \frac{g_{\phi\phi}}{g_{00}} g_{0\phi,\theta} \right) \Big] \partial_r \Bigg].\label{eq:32}
\end{align}
Outside the ergosphere (taking the angular velocity \( \Omega = 0 \)), \( \Omega_p = \Omega_{\text{LT}} \), and the above expression can be further simplified in polar coordinates as:
\begin{align}
	\Omega_p \big|_{\Omega=0} = \Omega_{\text{LT}} = \frac{1}{2\sqrt{-g}} \Big[ -\sqrt{g_{rr}} \left( g_{0\phi,\theta} - \frac{g_{0\phi}}{g_{00}} g_{00,\theta} \right) \hat{r} \notag \\
	+ \sqrt{g_{\theta\theta}} \left( g_{0\phi,r} - \frac{g_{0\phi}}{g_{00}} g_{00,r} \right) \hat{\theta} \Big].\label{eq:33}
\end{align}
For the complete expression of the Lense-Thirring precession frequency, refer to Appendix ~\ref{app:precession}.Inside the energy shell, the value of angular velocity \(\Omega\) must satisfy the condition that the Killing vector remains timelike: by requiring the Killing vector field to be timelike in this region (i.e., its norm is negative), the allowable range of \(\Omega\) can be derived as:

\begin{align}
	K^2 &= g_{tt} + 2\Omega g_{t\phi} + \Omega^2 g_{\phi\phi} < 0.\label{eq:34}
\end{align}

For convenience, we denote the angular velocity inside the energy shell as \(\Omega_e\). From equation (3.15), we obtain the following range for \(\Omega\):

\begin{align}
	\Omega_- &< \Omega_e < \Omega_+.\label{eq:35}
\end{align}

That is,

\begin{align}
	\Omega_{\pm} &= \frac{-g_{t\phi} \pm \sqrt{g_{t\phi}^2 - g_{tt} g_{\phi\phi}}}{g_{\phi\phi}}.\label{eq:36}
\end{align}

Substituting the corresponding metric components, we get:

\begin{align}
	\Omega_{\pm} &= \frac{
		-a t \sin^2 \theta \pm \sqrt{
			a^2 t^2 \sin^4 \theta - \sin^2 \theta 
		} 
	}{
		\sin^2 \theta \left( (r^2 + a^2)^2 - a^2 \Delta \sin^2 \theta \right)
	}\notag \\
	&\quad \times \frac{
		\sqrt{
			\left( (r^2 + a^2)^2 - a^2 \Delta \sin^2 \theta \right)(\Sigma^2 - t)
		}
	}{1}.\label{eq:37}
\end{align}

To find a suitable \(\Omega_e\), we introduce a parameter \(k\) (where \(0 < k < 1\)) to simplify Eq.\label{eq:37}\cite{rizwan2019distinguishing}:

\begin{align}
	\Omega_e &= k \Omega_+ + (1-k) \Omega_-.\label{eq:38}
\end{align}

That is,

\begin{align}
	\begin{split}
		\Omega_e &= \frac{
			-a t \sin^2 \theta + (2k - 1) \sqrt{
				\begin{aligned}[t]
					4a^2 t^2 \sin^4 \theta &- \sin^2 \theta \left( (r^2 + a^2)^2 \right. \\
					&\left. - a^2 \Delta \sin^2 \theta \right)(\Sigma^2 - t)
				\end{aligned}
			}
		}{
			\sin^2 \theta \left( (r^2 + a^2)^2 - a^2 \Delta \sin^2 \theta \right)
		} 
	\end{split}.\label{eq:39}
\end{align}

Substituting \(\Omega_e\) into Eq.\label{eq:32}, we obtain the complete expression for \(\Omega_p\),as shown in Appendix.~\ref{app:precession}

\subsection{Lense-Thirring (LT) Precession Frequency}
We now utilize eq.~\ref{eq:33} calculated in the previous section to examine the variation trends of the precession frequency under two dark matter models.
\begin{figure}[htbp]
	\centering
	\includegraphics[width=0.5\textwidth, keepaspectratio]{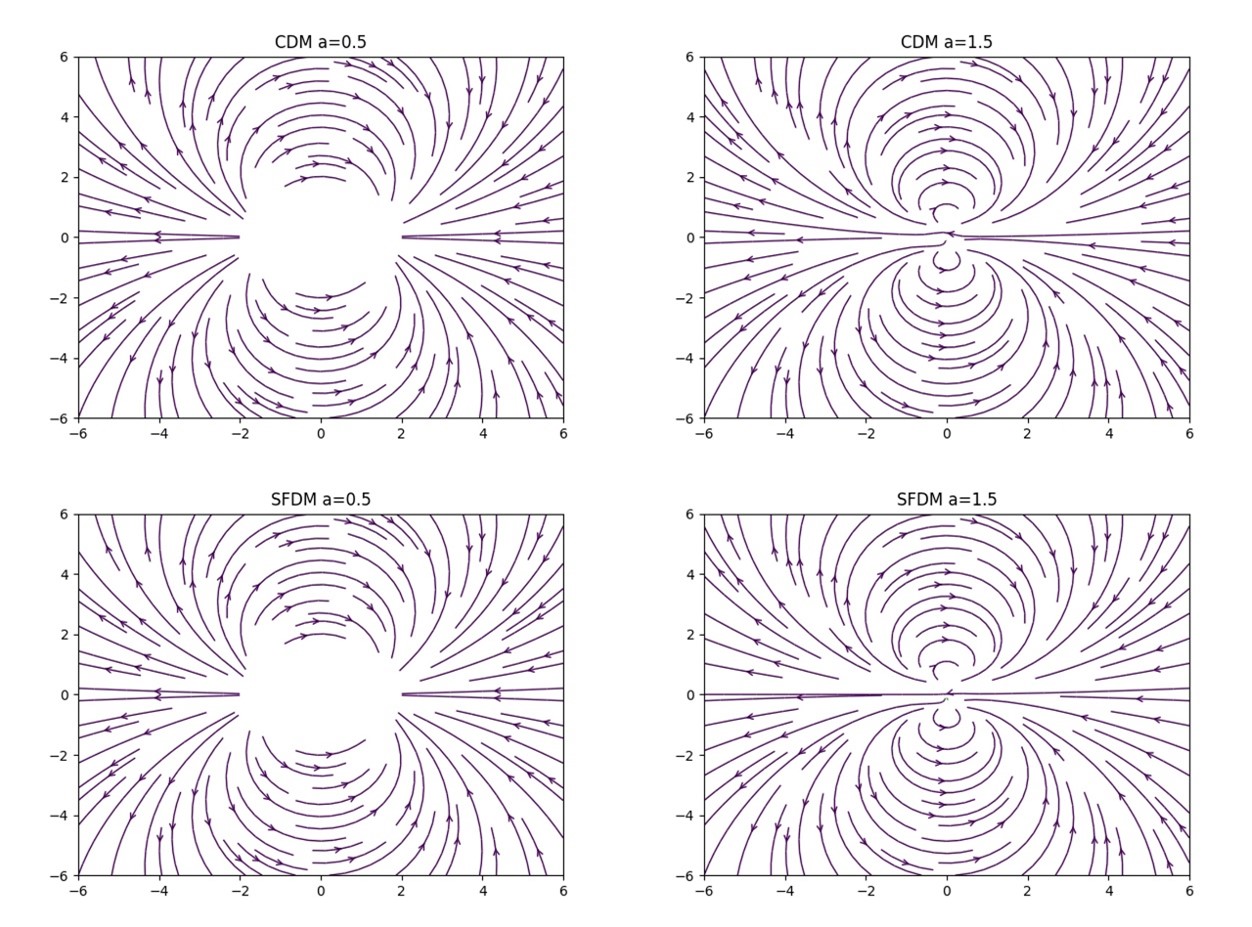}
	\caption{The vector plots on the Cartesian plane depict the Lense-Thirring (LT) precession frequency distribution, with parameters \( a = 0.5 \) characterizing the black hole configuration and \( a = 1.5 \) indicating the naked singularity scenario.}
	\label{fig:3}
\end{figure}
For the same dark matter model, when \(a = 0.5\), the precession frequency exists only outside the outer horizon, whereas when \(a = 1.5\), the precession frequency exists everywhere except at the ring singularity. Next, we discuss the numerical variation of the Lense-Thirring (LT) precession frequency. Numerically, the magnitude of the LT precession frequency is given by:
\begin{align}
	\Omega_{\text{LT}} = \frac{
		\sqrt{
			\begin{aligned}[t]
				& \left( -\sqrt{\Delta} \, 4at \cos\theta \sin\theta \left[ (r^2 + a^2)(\Sigma - t) + a^2 t \sin^2\theta \right] \right)^2 \\
				& + \left( 2a \Delta \sin^2\theta \Sigma (t' \Sigma - 2rt) \right)^2
			\end{aligned}
		}
	}{
		\left| 2 \Delta \Sigma^{7/2} \sin\theta \right|
	}.\label{eq:40}
\end{align}
We plotted the dependence of $\Omega_{\text{LT}}$ on $r$ for the CDM model with $a$ and $\theta$ as parameters, respectively.
\begin{figure}[htbp]
	\centering
	\begin{minipage}[b]{0.5\textwidth}
		\includegraphics[width=\textwidth, height=5cm, keepaspectratio]{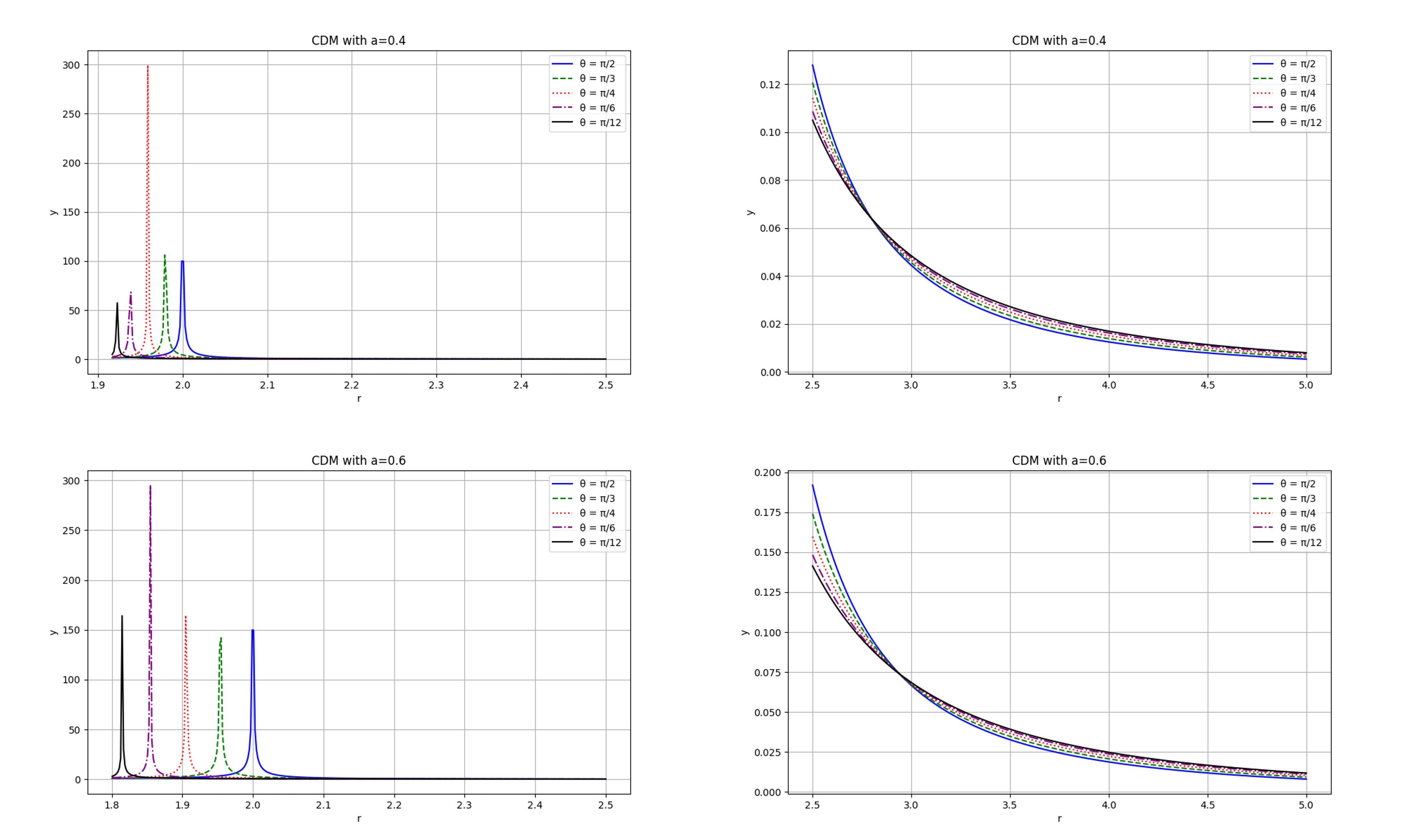}
	\end{minipage}
	\hfill
	\begin{minipage}[b]{0.5\textwidth}
		\includegraphics[width=\textwidth, height=5cm, keepaspectratio]{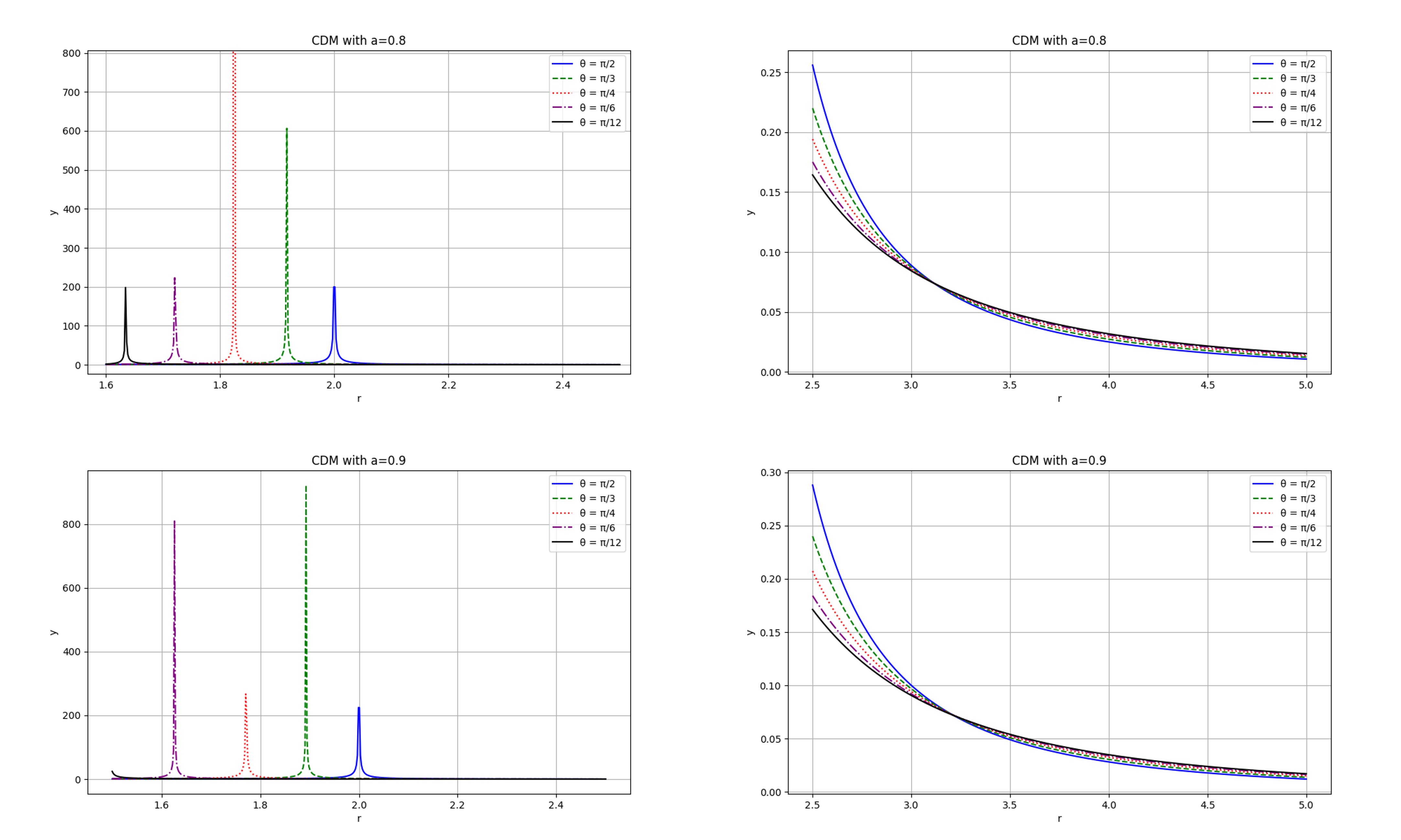}
	\end{minipage}
	\caption{The dependence of the LT precession frequency $\Omega_{\text{LT}}$ on $r$ in the CDM scenario is plotted, where the vertical axis $y$ represents $\Omega_{\text{LT}}$ and the horizontal axis denotes $r$.}\label{fig:4}
\end{figure}
A comparison between the curve intersections with the horizontal axis in Fig.~2 and Fig.~4 reveals that:
\begin{itemize}
	\item For black hole (BH) singularities ($a < 1$), the LT precession frequency diverges at the static limit surface (manifested as peaks in the figures due to finite computational accuracy)
	
	\item For naked singularity (NS) cases ($a > 1$), the divergence occurs where the metric component $g_{tt} = 0$, where the number of divergences coincides with the number of zero solutions of $g_{tt}$
\end{itemize}
We extend the same approach to analyze the SFDM model:
\begin{figure}[ht]
	\centering
	\begin{minipage}[b]{0.5\textwidth}
		\includegraphics[width=\textwidth, height=5cm, keepaspectratio]{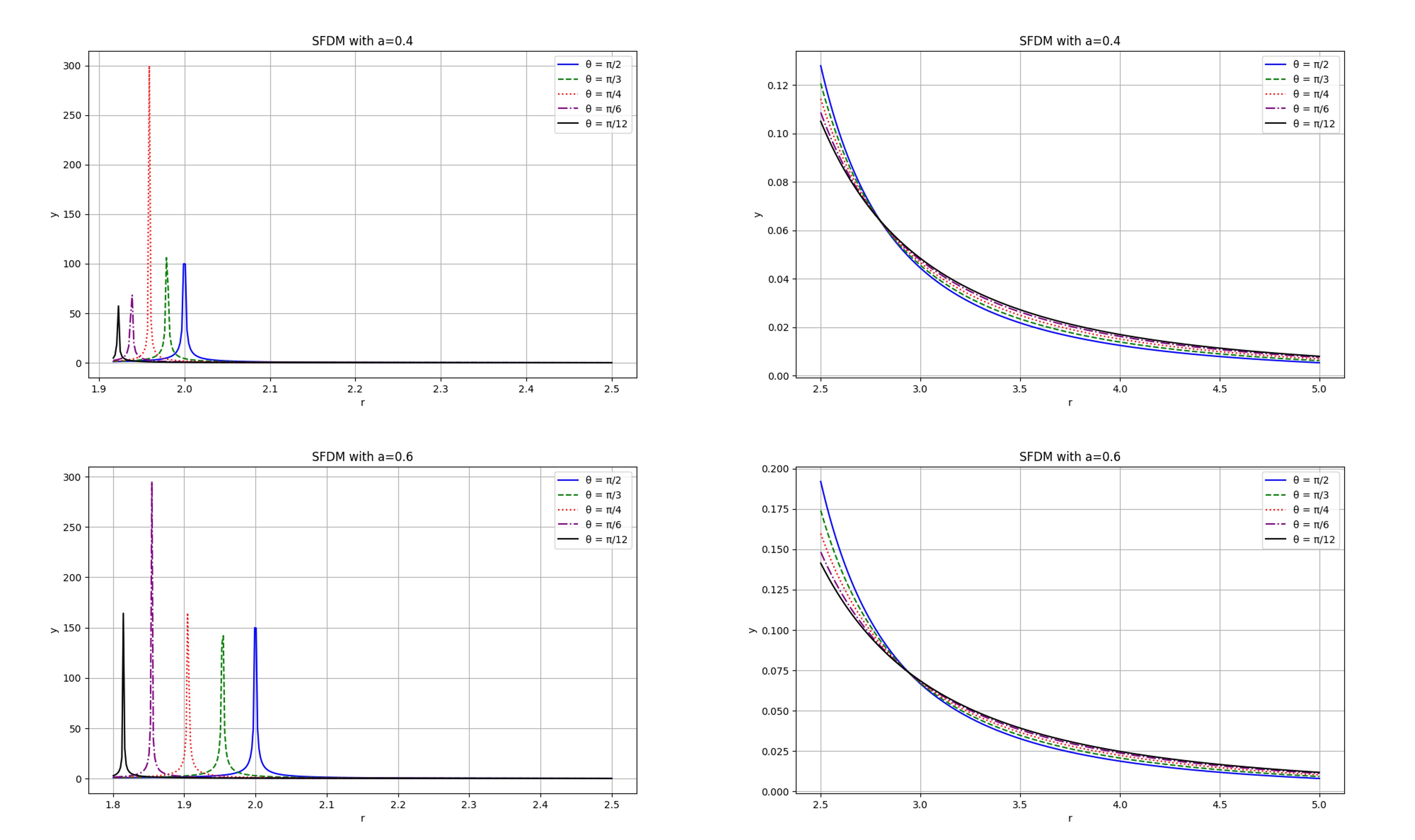}
	\end{minipage}
	\hfill
	\begin{minipage}[b]{0.5\textwidth}
		\includegraphics[width=\textwidth, height=5cm, keepaspectratio]{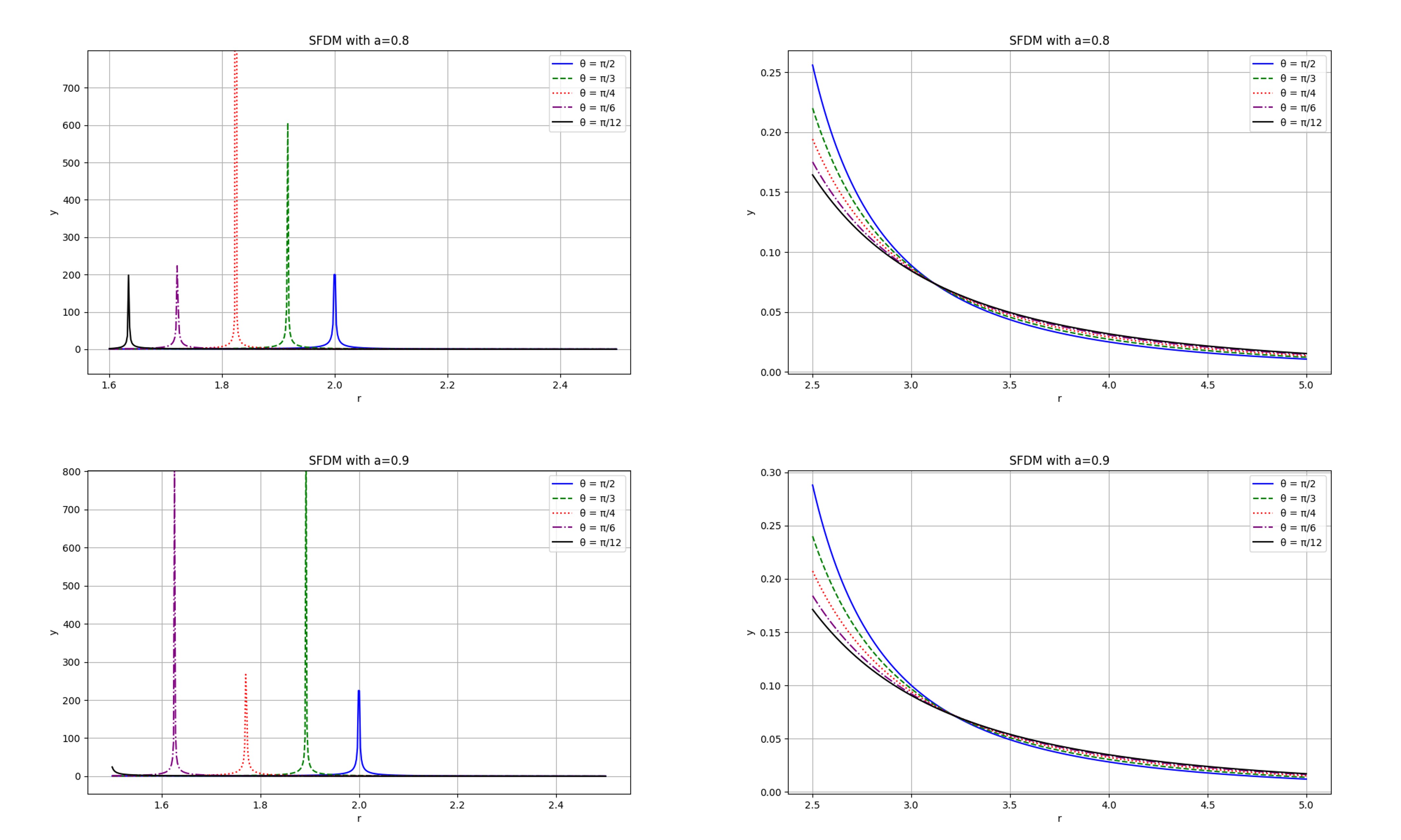}
	\end{minipage}
	\caption{The dependence of the LT precession frequency $\Omega_{\text{LT}}$ on $r$ in the CDM scenario is plotted, where the vertical axis $y$ represents $\Omega_{\text{LT}}$ and the horizontal axis denotes $r$.}\label{fig:5}
\end{figure}
It is not difficult to find by comparing Fig.~\ref{fig:5} with Fig.~\ref{fig:2} that, under scalar field dark matter (SFDM), when the singularity is a BH ($a<1$), the LT precession frequency $\Omega_{\text{LT}}$ diverges at the static limit surface (manifested in the figure as the form of peaks, which is limited by computational accuracy), while when the singularity is an NS ($a>1$), the LT precession frequency diverges at the point where the metric component $g_{tt}=0$, and the number of divergences is the same as the number of zero solutions of $g_{tt}$. 
From the expression of the LT precession frequency, we can easily find the relationship between the divergence of the precession frequency and $g_{tt}$. Specifically, it is not difficult to find that the expression of the LT precession frequency can be equivalently transformed into a form where the denominator is $g_{tt}$, so when $g_{tt}=0$, the LT precession frequency diverges. 
To clarify the influence of the cold dark matter model (CDM) and the scalar field dark matter model (SFDM) on LT precession, we set the dark matter parameters $\rho_c$ and $R_s$ both to 0 to obtain the precession frequency in Kerr spacetime, and then subtract the LT precession frequency under the cold dark matter model (CDM) and the scalar field dark matter model (SFDM) from the precession frequency in Kerr spacetime, respectively, to obtain $\Delta\Omega_{\text{LT}}$, as shown in the fig6 below:\\
\begin{figure}[ht]
	\centering
	\begin{subfigure}[b]{0.32\textwidth}
		\includegraphics[width=\textwidth, height=5cm, keepaspectratio]{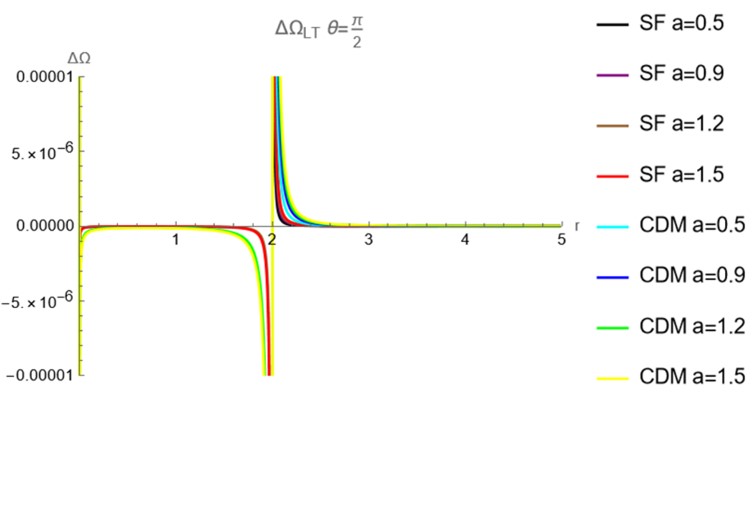}
		\caption{}
		\label{fig:6a}
	\end{subfigure}
	\hfill
	\begin{subfigure}[b]{0.32\textwidth}
		\includegraphics[width=\textwidth, height=5cm, keepaspectratio]{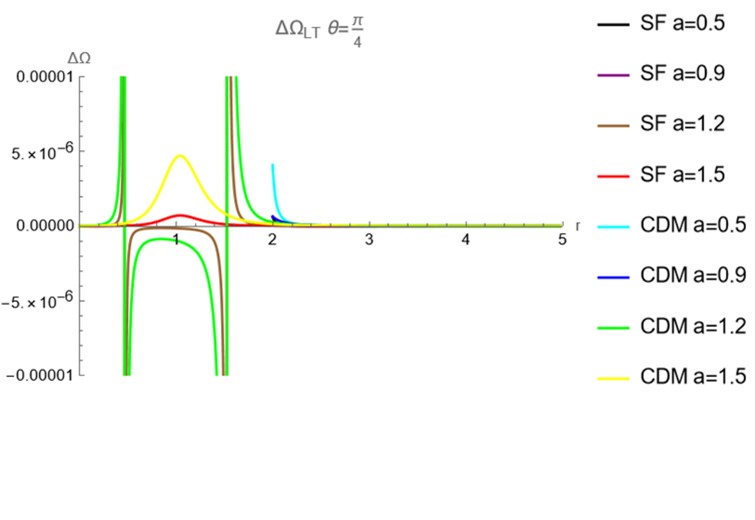}
		\caption{}
		\label{fig:6b}
	\end{subfigure}
	\hfill
	\begin{subfigure}[b]{0.32\textwidth}
		\includegraphics[width=\textwidth, height=5cm, keepaspectratio]{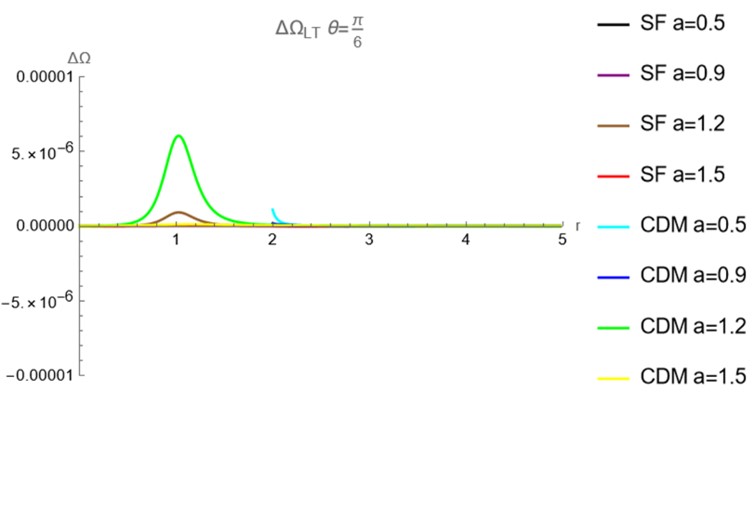}
		\caption{}
		\label{fig:6c}
	\end{subfigure}
	
	\caption{Comparison of frame-dragging frequency difference $\Delta\Omega_{\text{LT}}$ between CDM/SFDM models and Kerr spacetime. Spin parameters $a$ values are annotated in legend. Dashed vertical lines indicate divergence positions at $g_{tt} = 0$.}
	\label{fig:6}
\end{figure}
From the  Fig.~\ref{fig:6}, it is evident that for black hole (BH) singularities, both CDM and SFDM cause divergence in the LT precession frequency at the static limit surface, with CDM having a significantly stronger influence than SFDM. For naked singularities (NS) in non-equatorial planes, apart from the divergence at \( g_{tt} = 0 \), CDM and SFDM exhibit an additional peak in their effects on the LT precession frequency, as shown by the green (CDM \( a=1.2 \)), yellow (CDM \( a=1.5 \)), brown (SFDM \( a=1.2 \)), and red (SFDM \( a=1.5 \)) curves in panels (b) and (c). Furthermore, the peak amplitude (within the same dark matter model) decreases as the spin parameter \( a \) increases. Notably, the numerical values of the peaks indicate that cold dark matter has a significantly greater impact on LT precession compared to scalar field dark matter.
\subsection{The Precession Frequency Within The Ergoregion}
In the previous section, we discussed the precession behavior of gyroscopes outside the ergoregion. In this section, we will analyze the precession of gyroscopes in the equatorial plane under general conditions (including both inside and outside the ergoregion). From Equation (3.17), we derive the variation of the precession frequency within the ergoregion as a function of \( r \). We categorize the observations into three cases based on the parameter \( q \): \( q < 0.5 \), \( q = 0.5 \), and \( q > 0.5 \)\\
\begin{figure}[h]
	\centering
	\includegraphics[width=0.5\textwidth, keepaspectratio]{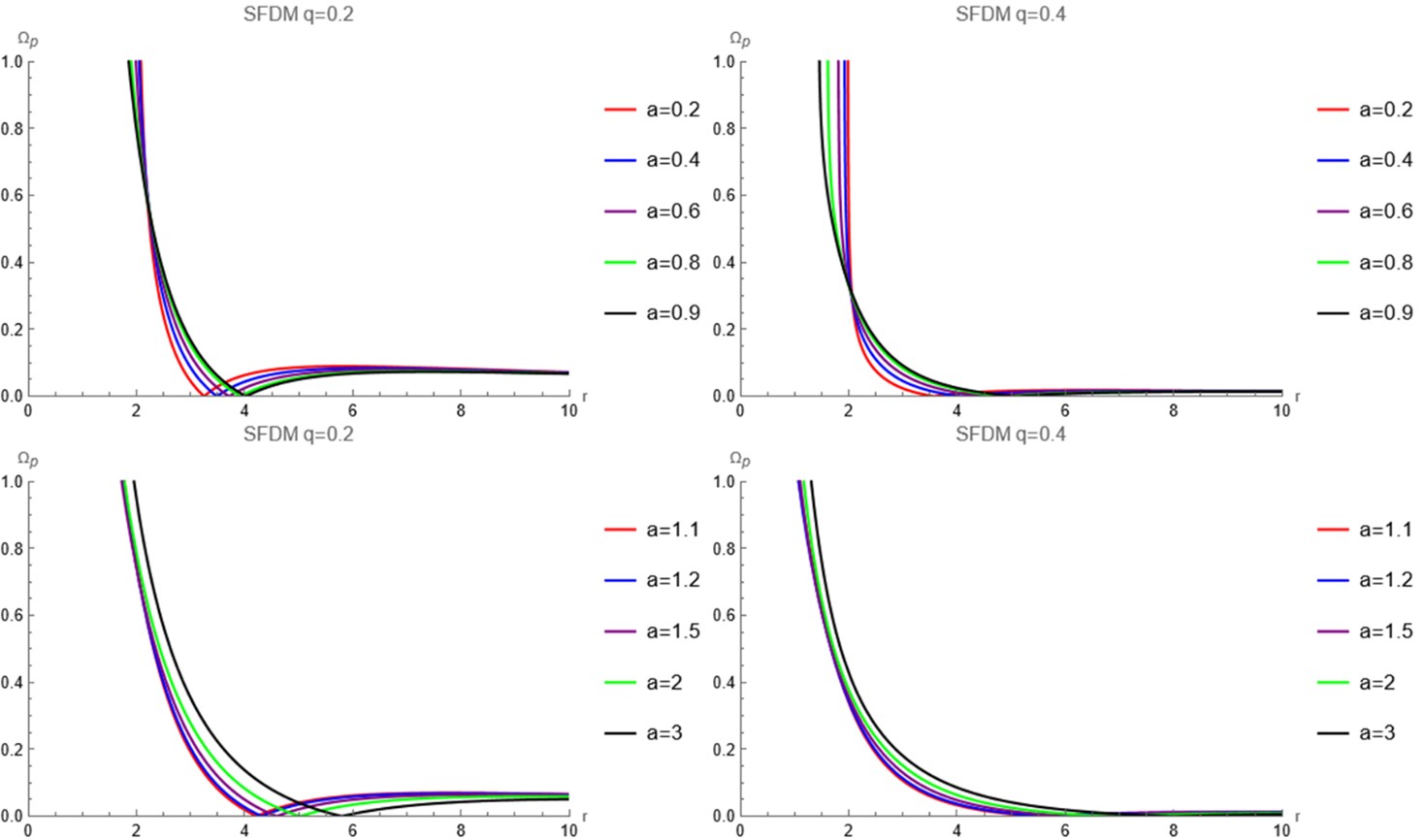}
	\caption{The trend of the precession frequency with respect to \( r \) in the SFDM case at \( \theta = \pi/2 \) is plotted. We select the parameter \( q \) as 0.2 and 0.4, with the angle \( \theta = \pi/2 \), and discuss the variation of the precession frequency under different \( q \) values for the cases where the singularity is a BH (black hole spin parameter \( a \) as 0.2, 0.4, 0.6, 0.8, 0.9) and where the singularity is an NS (black hole spin parameter \( a \) as 1.1, 1.2, 1.5, 2, 3).}
	\label{fig:7}
\end{figure}
From the Fig.~\ref{fig:7}, we notice that the position of the trough increases with the increase of the spin parameter \( a \) under the same parameter \( q \); under the same spin parameter \( a \), it increases with the increase of the parameter \( q \).
\begin{figure}[h]
	\centering
	\includegraphics[width=0.5\textwidth, keepaspectratio]{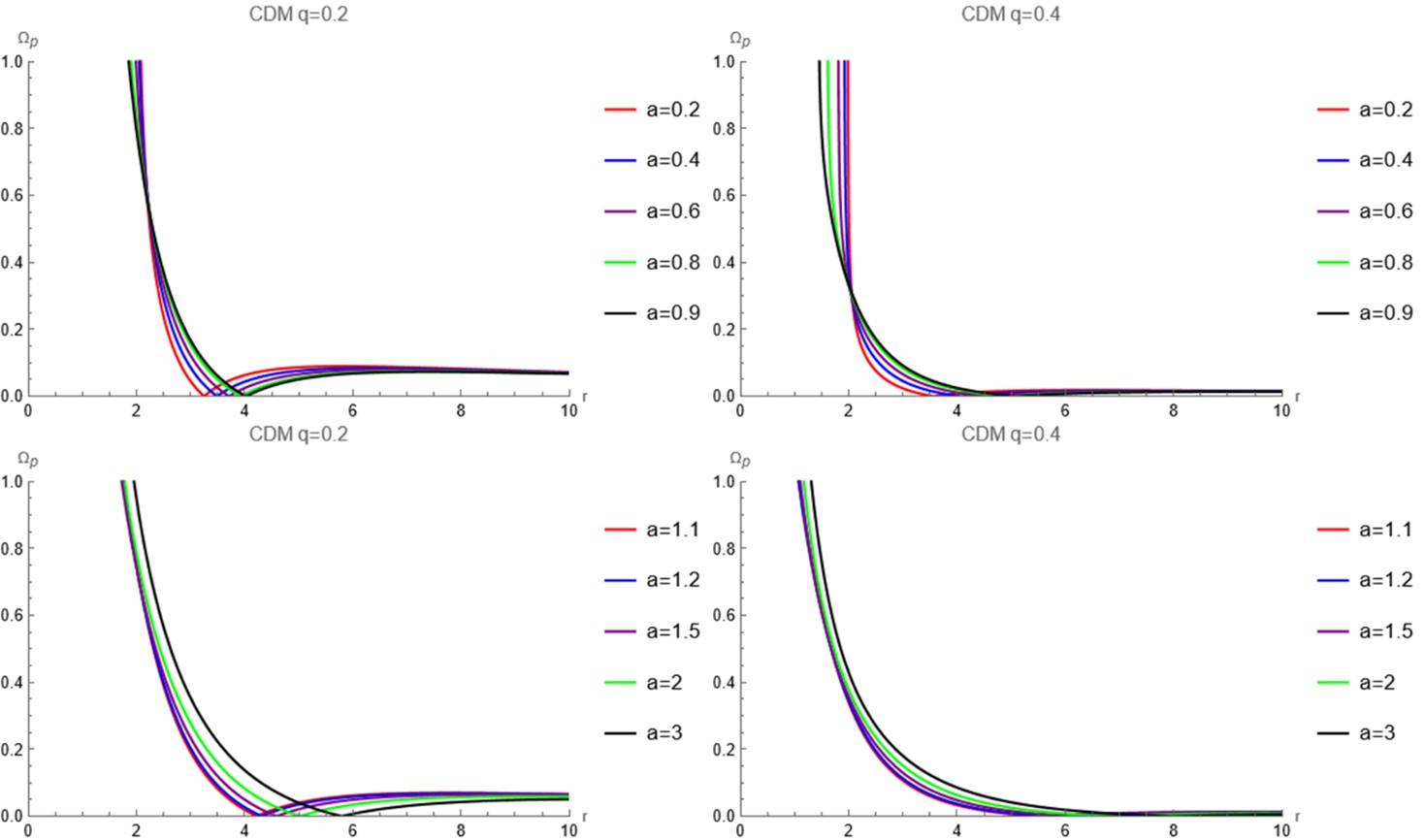}
	\caption{The variation trend of the precession frequency with respect to \( r \) in the CDM case is plotted. Similarly, we select the parameter \( q \) as 0.2 and 0.4, with the angle \( \theta = \pi/2 \), and discuss the variation of the precession frequency under different \( q \) values for the cases where the singularity is a BH (black hole spin parameter \( a \) as 0.2, 0.4, 0.6, 0.8, 0.9) and where the singularity is an NS (black hole spin parameter \( a \) as 1.1, 1.2, 1.5, 2, 3).}
	\label{fig:8}
\end{figure}
In Fig.~\ref{fig:8}, the position where the precession frequency exhibits a trough increases with the increase of the spin parameter \( a \) under the same parameter \( q \); under the same spin parameter \( a \), it increases with the increase of the parameter \( q \).
\begin{figure}[h]
	\centering
	\includegraphics[width=0.5\textwidth, keepaspectratio]{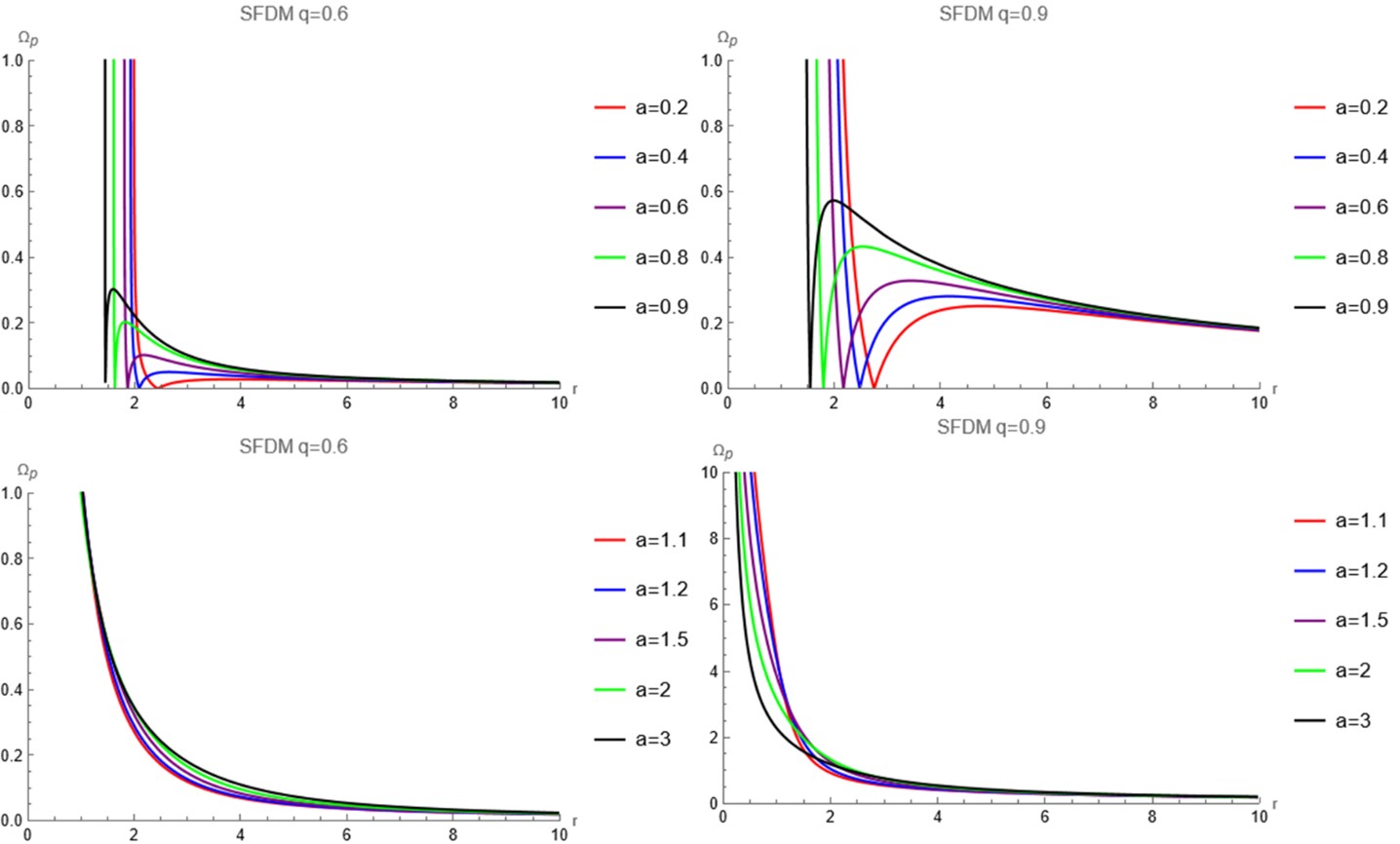}
	\caption{The figure depicts the variation trend of the precession frequency with respect to \( r \) in the SFDM case. We select the parameter \( q \) as 0.6 and 0.9, with the angle \( \theta = \pi/2 \), and discuss the variation of the precession frequency under different \( q \) values for the cases where the singularity is a BH (black hole spin parameter \( a \) as 0.2, 0.4, 0.6, 0.8, 0.9) and where the singularity is an NS (black hole spin parameter \( a \) as 1.1, 1.2, 1.5, 2, 3).}
	\label{fig:9}
\end{figure}
In the Fig.~\ref{fig:9}, when the singularity is a BH, the precession frequency exhibits a fluctuation, a feature absent when the singularity is a naked singularity. The peak amplitude decreases with the increase of the spin parameter \( a \) and increases with the increase of the parameter \( q \).
\begin{figure}[h]
	\centering
	\includegraphics[width=0.5\textwidth, keepaspectratio]{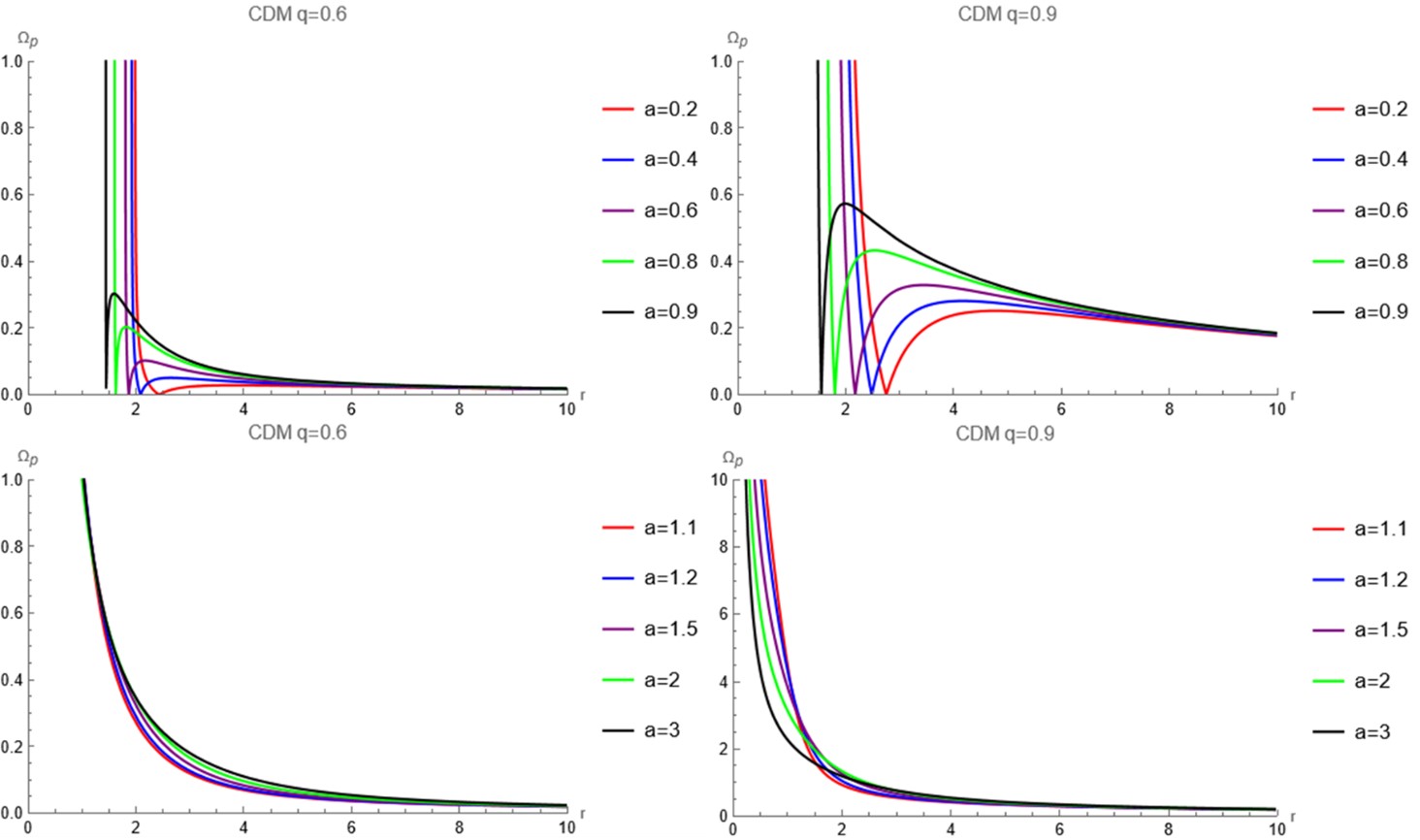}
	\caption{The figure illustrates the variation trend of the precession frequency with respect to \( r \) in the CDM case. We select the parameter \( q \) as 0.6 and 0.9, with the angle \( \theta = \pi/2 \), and discuss the variation of the precession frequency under different \( q \) values for the cases where the singularity is a BH (black hole spin parameter \( a \) as 0.2, 0.4, 0.6, 0.8, 0.9) and where the singularity is an NS (black hole spin parameter \( a \) as 1.1, 1.2, 1.5, 2, 3).}
	\label{fig:10}
\end{figure}
In Fig.~\ref{fig:10}, under the cold dark matter (CDM) scenario, the precession frequency shows a pronounced fluctuation when the singularity is a black hole (BH), a feature absent when it is a naked singularity (NS).
\begin{figure}[h]
	\centering
	\includegraphics[width=0.5\textwidth, keepaspectratio]{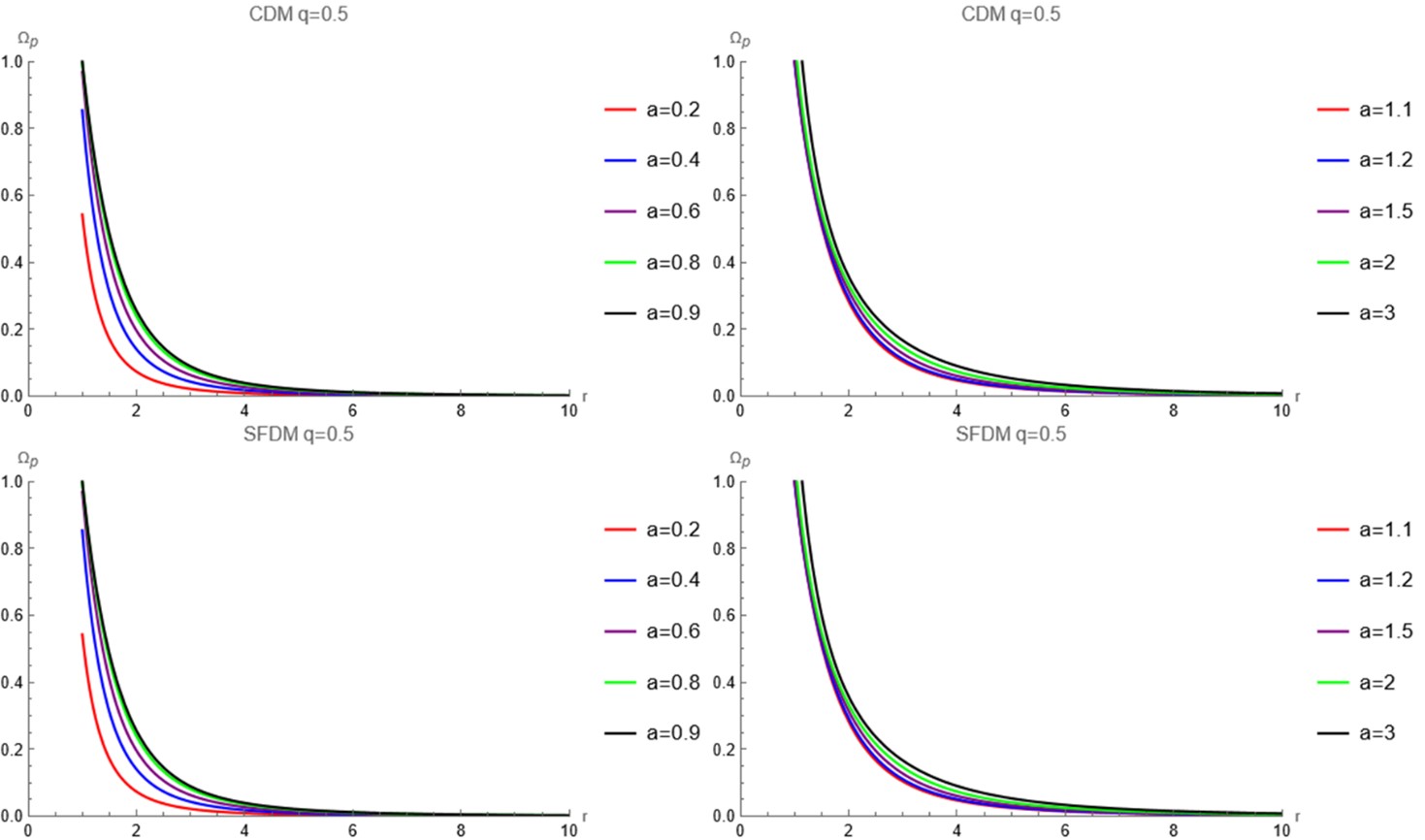}
	\caption{The precession frequency in the equatorial plane as a function of \( r \) is plotted for \( q = 0.5 \) in both the CDM and SFDM scenarios. When the singularity is a BH, the black hole spin parameter \( a \) is taken as 0.2, 0.4, 0.6, 0.8, 0.9; when the singularity is an NS, the black hole spin parameter \( a \) is taken as 1.1, 1.2, 1.5, 2, 3. }
	\label{fig:11}
\end{figure}
In Fig.~\ref{fig:11}, the precession frequency decreases with increasing \( r \) for both BH and NS singularities. At the same position (i.e., the same \( r \)), the precession frequency increases with increasing spin parameter \( a \). When the singularity is a BH, the precession frequency diverges at the horizon; when the singularity is an NS, it similarly diverges near the singularity.
\section{\label{sec:level4}Dynamics of the Innermost Stable Circular Orbit in Black Hole Accretion Disks and Characteristic Frequency Analysis under Dark Matter Models}
This section is dedicated to studying the frame-dragging effect of black hole accretion disks by analyzing the dynamical properties of the innermost stable circular orbit (ISCO) to establish key distinguishing criteria. Based on the four-velocity normalization condition 
\begin{align}
	-1 = g_{\mu\nu} \dot{x}^\mu \dot{x}^\nu \quad (\text{where } \dot{x}^\mu \equiv dx^\mu / d\tau).\label{eq:41}
\end{align}
the conserved quantities are derived as follows:
\begin{align}
	L &= g_{t\phi} \frac{dt}{d\tau} + g_{\phi\phi} \frac{d\phi}{d\tau}.\label{eq:42} \quad \\
	E &= -g_{tt} \frac{dt}{d\tau} - g_{t\phi} \frac{d\phi}{d\tau}.\label{eq:43} \quad 
\end{align}

By combining Eq.\eqref{eq:43} and Eq.\eqref{eq:44} with the normalization condition, the evolution rates of the time and angular components are obtained:
\begin{align}
	\frac{dt}{d\tau} &= \frac{E g_{\phi\phi} + L g_{t\phi}}{g_{t\phi}^2 - g_{tt} g_{\phi\phi}}.\label{eq:44} \\
	\frac{d\phi}{d\tau} &= \frac{E g_{t\phi} + L g_{tt}}{g_{tt} g_{\phi\phi} - g_{t\phi}^2}.\label{eq:45}
\end{align}

The radial motion equation is determined by the metric components:
\begin{align}
	\frac{dr}{d\tau} &= \sqrt{\frac{-1 - \left[ g_{tt} \left( \frac{dt}{d\tau} \right)^2 + 2 g_{t\phi} \frac{dt}{d\tau} \frac{d\phi}{d\tau} + g_{\phi\phi} \left( \frac{d\phi}{d\tau} \right)^2 \right]}{g_{rr}}}.\label{eq:46}
\end{align}

After reorganization, the geodesic motion equation in the equatorial plane is:
\begin{align}
	\frac{E^2 - 1}{2} &= \left( \frac{dr}{d\tau} \right)^2 + V_{\text{eff}}(r), \label{eq:47}
\end{align}
where \( V_{\text{eff}}(r) \) is the effective potential function, and \( (E^2 - 1)/2 \) represents the total mechanical energy of the particle. The existence of stable orbits requires the simultaneous satisfaction of the extremum conditions:
\begin{align}
	\frac{d V_{\text{eff}}}{dr} &= 0,\label{eq:48} \\
	\frac{d^2 V_{\text{eff}}}{dr^2} &\geq 0.\label{eq:49}
\end{align}

The ISCO radius is identified as the first critical point satisfying these conditions. By analyzing the regulatory effects of the spin parameter \( a \) and dark matter halo parameters on the second derivative of \( V_{\text{eff}}(r) \), a criterion system can be constructed to distinguish between black holes and naked singularities, as well as between CDM and SFDM models.
\begin{table}[htbp]
	\centering
	\begin{tabular}{S[table-format=1.1] S[table-format=1.2] c}
		\toprule
		{$a$ (\si{/M})} & {$r|_{v'=0}$ (\si{/M})} & {Sign of $v''$} \\
		\midrule
		0.2 & 3.94 & + \\
		0.4 & 3.49 & + \\
		0.5 & 3.24 & + \\
		0.6 & 2.97 & + \\
		0.8 & 2.38 & + \\
		1.2 & 1.12 & + \\
		1.5 & 1.63 & + \\
		0.0 & 6.00 & +\\
		\bottomrule
	\end{tabular}
	\caption{Numerical analysis in the CDM scenario using the effective potential derivatives $v'=0$ and $v''=0$. The symbols "+" and "$-$" denote the sign of $v''$ at $v''=0$ being positive (+) or negative ($-$).}
	\label{tab:cdm}
\end{table}

\begin{table}[htbp]
	\centering
	\begin{tabular}{S[table-format=1.1] S[table-format=1.2] c}
		\toprule
		{$a$ (\si{/M})} & {$r|_{v'=0}$ (\si{/M})} & {Sign of $v''$} \\
		\midrule
		0.2 & 3.95 & + \\
		0.4 & 3.51 & + \\
		0.5 & 3.25 & + \\
		0.6 & 2.98 & + \\
		0.8 & 2.39 & + \\
		1.2 & 1.13 & + \\
		1.5 & 1.64 & + \\
		0.0 & 6.00 & +\\
		\bottomrule
	\end{tabular}
	\caption{SFDM scenario, with the same symbol convention as in Table~\ref{tab:cdm}.}
	\label{tab:sfdm}
\end{table}

\begin{table}[htbp]
	\centering
	\begin{tabular}{S[table-format=1.1] S[table-format=1.2] c}
		\toprule
		{$a$ (\si{/M})} & {$r|_{v'=0}$ (\si{/M})} & {Sign of $v''$} \\
		\midrule
		0.2 & 5.33 & + \\
		0.4 & 4.61 & + \\
		0.5 & 4.23 & + \\
		0.6 & 3.83 & + \\
		0.8 & 2.91 & + \\
		1.2 & 0.70 & + \\
		1.5 & 0.88 & + \\
		0.0 & 6.00 & +\\
		\bottomrule
	\end{tabular}
	\caption{Kerr scenario with dark matter parameters $\rho_c$ and $R_s$ set to zero, reverting to the Kerr case \cite{chakraborty2017distinguishing}}
	\label{tab:kerr}
\end{table}

From Tables~\ref{tab:cdm} and~\ref{tab:sfdm}, it can be observed that the calculated \(r|_{v'=0}\) lies to the left of \(r|_{v''=0}\), and \(v''\) is a decreasing function. Therefore, the \(r|_{v'=0}\) values computed in Tables~\ref{tab:cdm} and~\ref{tab:sfdm} correspond to the radius of the innermost stable circular orbit (ISCO). Additionally, setting the spin parameter \(a\) and dark matter parameters \(\rho_c\) and \(R_s\) to zero successfully recovers the Schwarzschild case. Comparing Tables~\ref{tab:cdm} and~\ref{tab:sfdm}, the choice of dark matter type (CDM vs. SFDM) has a very weak effect on \(r_{\text{ISCO}}\). To determine whether this arises from the specific dark matter models or from a generally weak influence of dark matter, we computed \(r_{\text{ISCO}}\) in the Kerr metric without dark matter parameters (Table~\ref{tab:kerr}).

Based on the numerical results from Tables~\ref{tab:cdm} (CDM model),~\ref{tab:sfdm} (SFDM model), and~\ref{tab:kerr} (Kerr baseline), a systematic analysis of the effective potential’s first derivative (\(v'=0\)) and second derivative stability condition (\(v''>0\)) reveals: Although the ISCO radius difference between CDM and SFDM models is negligible within computational precision (\(|\Delta r_{\text{ISCO}}| < 10^{-4}\)), introducing dark matter halo parameters (\(\rho_c, R_s \neq 0\)) induces a systematic shift in \(r_{\text{ISCO}}\) of approximately \(10^{-2}\) order relative to the Kerr spacetime (e.g., at \(a=0\), \(r_{\text{ISCO}}\) decreases from 6.00 to 5.33–3.94 in CDM/SFDM). This demonstrates an observable correction to accretion dynamics due to dark matter’s gravitational potential. Notably, the spin parameter \(a\) exerts a significantly stronger influence on \(r_{\text{ISCO}}\) than dark matter model differences, with a variation gradient of \(\Delta r \sim 0.1\) (e.g., at \(a=1.5\), \(r_{\text{ISCO}}\) drops to 0.88–1.63). The positive \(v'' > 0\) across all models confirms Lyapunov stability under radial perturbations. This suggests that observing \(r_{\text{ISCO}}\) and its response to \(a\) can constrain dark matter presence. Furthermore, the characteristic frequency combinations (\(\Omega_r, \Omega_\theta, \Omega_\phi\)) excited by perturbations may probe dark matter types. For an object in circular motion at radius \(r_c\) under perturbation, it oscillates radially and vertically at frequencies \(\omega_r\) and \(\omega_\theta\). These, with the Keplerian frequency \(\Omega_\phi\), form the accretion disk’s fundamental frequencies: \(\Omega_\phi\) (Keplerian), \(\Omega_r\) (radial epicyclic), and \(\Omega_\theta\) (vertical epicyclic)\cite{doneva2014orbital,pappas2012can}.
The frequencies \(\Omega_r^2\), \(\Omega_\theta^2\), and \(\Omega_\phi\) are expressed as follows\cite{doneva2014orbital,ryan1995gravitational}:

\begin{align}
	\Omega_r^2 &= -\frac{1}{g_{rr}} \left[
	\begin{aligned}
		& (g_{tt} + g_{t\phi} \Omega_\phi)^2 \, \partial_r^2 \left( \frac{g_{\phi\phi}}{\rho^2} \right) \\
		& - 2 (g_{tt} + g_{t\phi} \Omega_\phi)(g_{t\phi} + g_{\phi\phi} \Omega_\phi) \, \partial_r^2 \left( \frac{g_{t\phi}}{\rho^2} \right) \\
		& + (g_{t\phi} + g_{\phi\phi} \Omega_\phi)^2 \, \partial_r^2 \left( \frac{g_{tt}}{\rho^2} \right)
	\end{aligned}
	\label{eq:50}
	\right]. \\
	\Omega_\theta^2 &= -\frac{1}{g_{rr}} \left[
	\begin{aligned}
		& (g_{tt} + g_{t\phi} \Omega_\phi)^2 \, \partial_\theta^2 \left( \frac{g_{\phi\phi}}{\rho^2} \right) \\
		& - 2 (g_{tt} + g_{t\phi} \Omega_\phi)(g_{t\phi} + g_{\phi\phi} \Omega_\phi) \, \partial_\theta^2 \left( \frac{g_{t\phi}}{\rho^2} \right) \\
		& + (g_{t\phi} + g_{\phi\phi} \Omega_\phi)^2 \, \partial_\theta^2 \left( \frac{g_{tt}}{\rho^2} \right)
	\end{aligned}
	\label{eq:51}
	\right].\\
	\Omega_\phi &= \frac{-g_{t\phi,r} \pm \sqrt{(g_{t\phi,r})^2 - g_{tt,r} g_{\phi\phi,r}}}{g_{\phi\phi,r}}.
	\label{eq:52}
\end{align}

where
\[
\rho^2 = g_{tt} g_{\phi\phi} - g_{t\phi}^2.
\]

These frequencies are related to the precessions of the orbit and the orbital plane. The orbital precession is represented by the periastron precession frequency \(\Omega_{\text{per}}\), and the precession of the orbital plane is represented by the nodal precession frequency \(\Omega_{\text{nod}}\)\cite{belloni2015fast}:
\begin{align}
	\Omega_{\text{per}} &= \Omega_\phi - \Omega_r, \label{eq:53}\\
	\Omega_{\text{nod}} &= \Omega_\phi - \Omega_\theta.\label{eq:54} \quad
\end{align}

Using \eqref{eq:50},\eqref{eq:51},\eqref{eq:52},\eqref{eq:53} and \eqref{eq:54}, the variations of \(\Omega_r^2\), \(\Omega_\theta^2\), \(\Omega_\phi\), \(\Omega_{\text{per}}\), and \(\Omega_{\text{nod}}\) with respect to \(r\) can be obtained.
\subsection{$\Omega_r^2$}
\begin{figure}[htbp]
	\centering
	\begin{minipage}[b]{0.5\textwidth}
		\includegraphics[width=\textwidth, height=5cm, keepaspectratio]{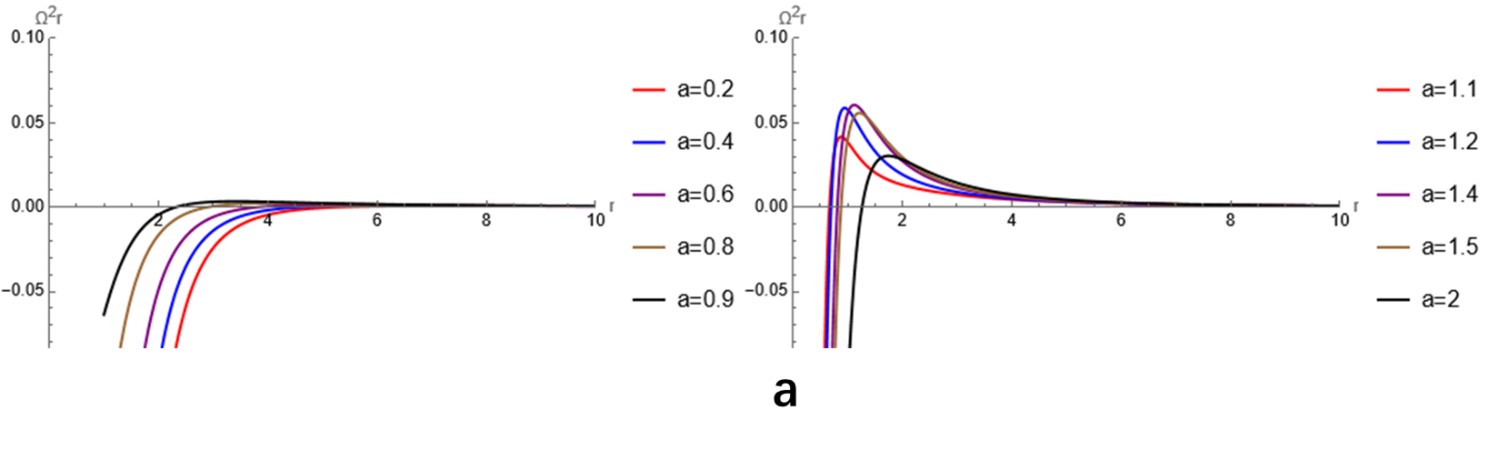}
	\end{minipage}
	\hfill
	\begin{minipage}[b]{0.5\textwidth}
		\includegraphics[width=\textwidth, height=5cm, keepaspectratio]{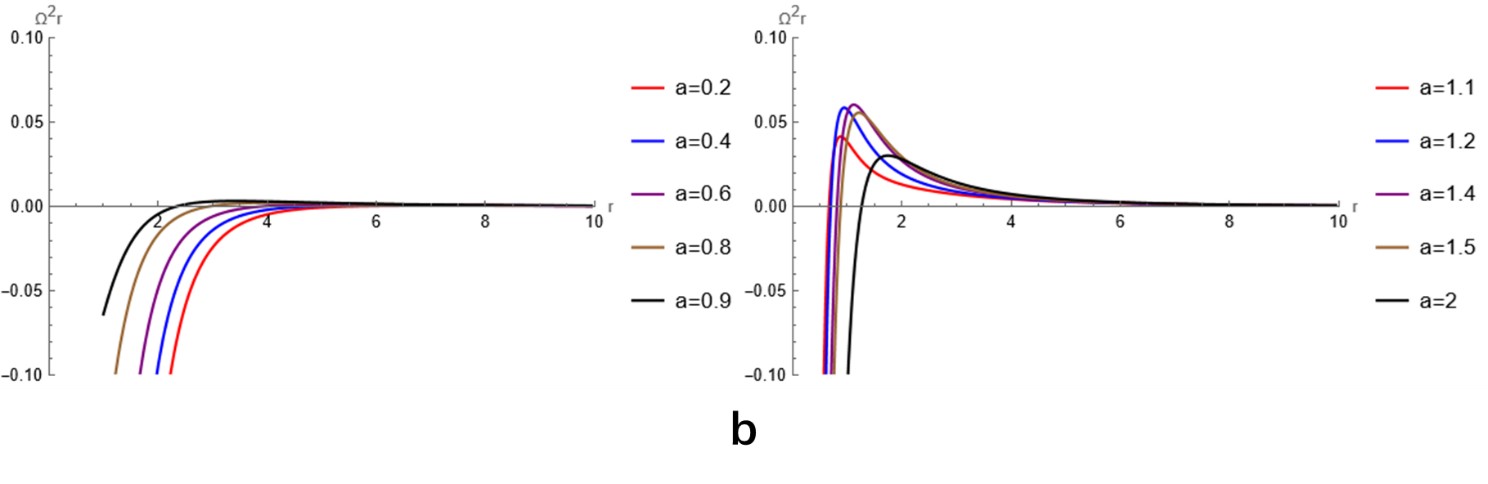}
	\end{minipage}
	\caption{The figure illustrates the variation of $\Omega_r^2$ with $r$ for the cold dark matter (CDM) case (a) and the scalar field dark matter (SFDM) case (b).}\label{fig:12}
\end{figure}
In Fig.~\ref{fig:12}, for both the CDM and SFDM cases, when the singularity is a black hole (BH) or a naked singularity (NS), the $\Omega_r^2$ curve exhibits a peak. In the BH case, the location of the peak ($r$) decreases as the spin parameter $a$ increases, with the peak always appearing to the right of $r = 2M$. Additionally, the height of the peak increases as the spin parameter $a$ increases. In the NS case, the height of the peak no longer shows a linear increase with $a$, and the location of the peak ($r$) increases as $a$ increases. The peak for NS always appears to the left of $r = 2M$, and the peak height for NS is significantly larger than that for BH.
\subsection{$\Omega_\theta^2$}
\begin{figure}[htbp]
	\centering
	\begin{minipage}[b]{0.5\textwidth}
		\includegraphics[width=\textwidth, height=5cm, keepaspectratio]{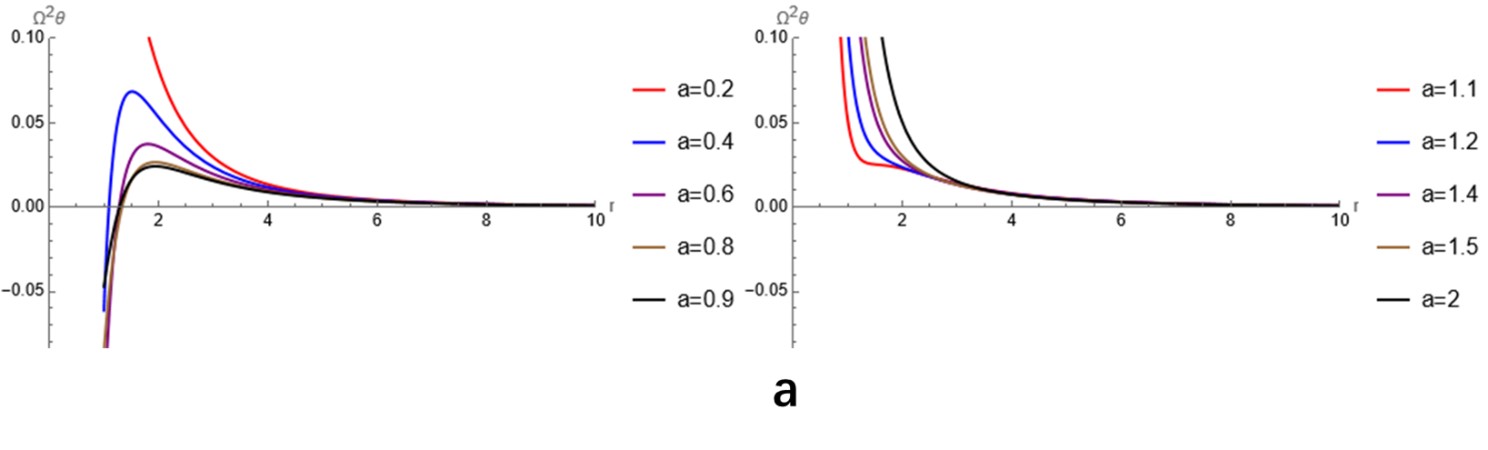}
	\end{minipage}
	\hfill
	\begin{minipage}[b]{0.5\textwidth}
		\includegraphics[width=\textwidth, height=5cm, keepaspectratio]{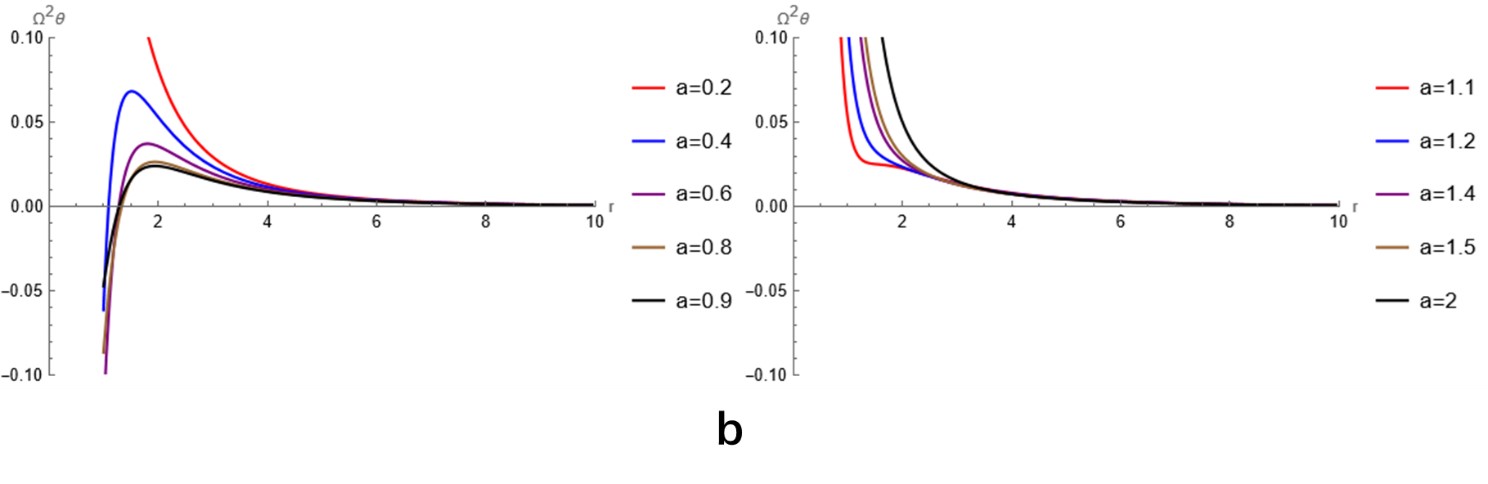}
	\end{minipage}
	\caption{The figure illustrates the variation of $\Omega_\theta^2$ with $r$ for cold dark matter (CDM) (a) and scalar field dark matter (SFDM) (b).}\label{fig:13}
\end{figure}
In Fig.~\ref{fig:13}, for the case of CDM and SFDM, when the singularity is a black hole (BH), the $\Omega_\theta^2$ curve exhibits a peak, and the peak value decreases as the spin parameter $a$ increases, with regions where $\Omega_\theta^2 < 0$ existing. When the singularity is a naked singularity (NS), $\Omega_\theta^2$ is always nonzero and decreases monotonically with increasing $r$.
\subsection{$\Omega_\phi$}
\begin{figure}[htbp]
	\centering
	\begin{minipage}[b]{0.5\textwidth}
		\includegraphics[width=\textwidth, height=5cm, keepaspectratio]{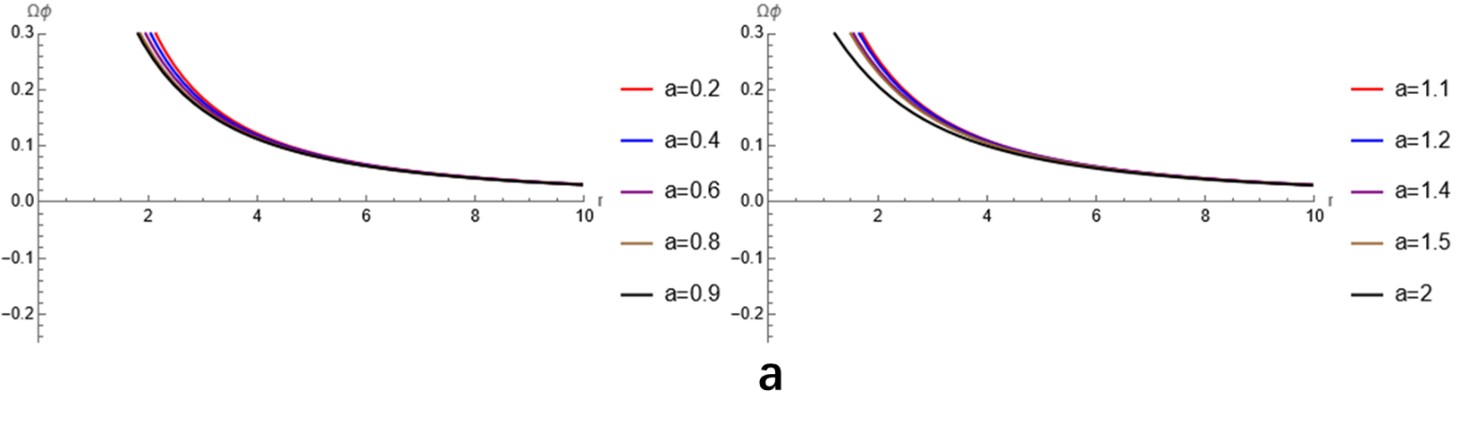}
	\end{minipage}
	\hfill
	\begin{minipage}[b]{0.5\textwidth}
		\includegraphics[width=\textwidth, height=5cm, keepaspectratio]{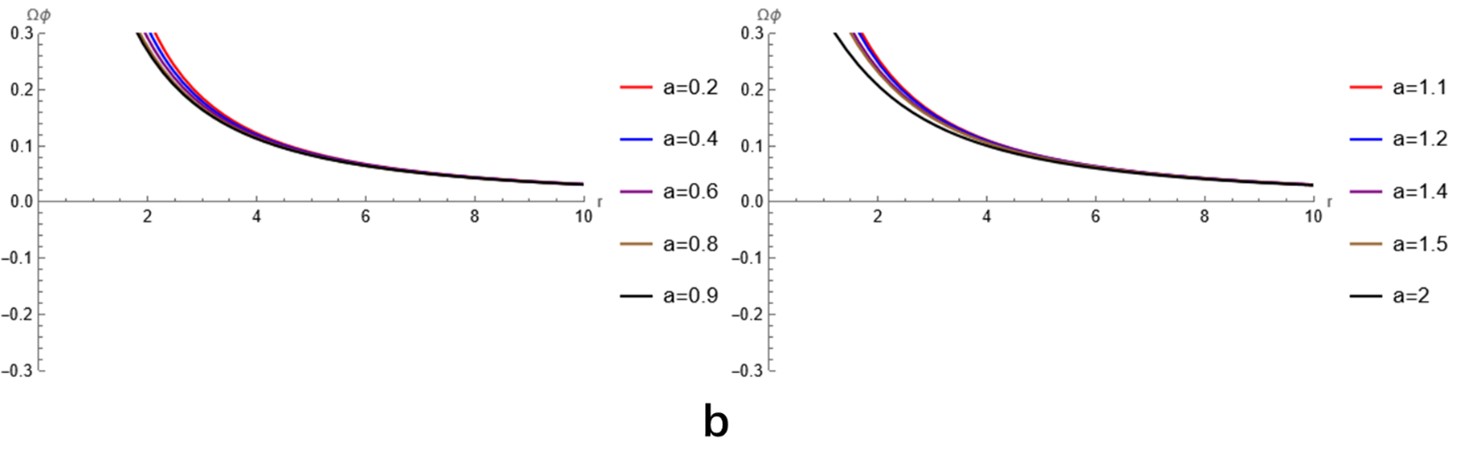}
	\end{minipage}
	\caption{The figure illustrates the variation of $\Omega_\phi$ with $r$ for cold dark matter (CDM) (a) and scalar field dark matter (SFDM) (b).}\label{fig:14}
\end{figure}
In Fig.~\ref{fig:14}, For both the CDM and SFDM cases, $\Omega_\phi$ decreases monotonically as $r$ increases. At the same radius $r$, $\Omega_\phi$ decreases as the spin parameter $a$ increases (whether $a < 1$ or $a > 1$). However, when $a = 0.9$, the $\Omega_\phi$ curve has lower values compared to the $\Omega_\phi$ curve when $a = 1.1$.
\subsection{$\Omega_{nod}$}
\begin{figure}[htbp]
	\centering
	\begin{minipage}[b]{0.5\textwidth}
		\includegraphics[width=\textwidth, height=5cm, keepaspectratio]{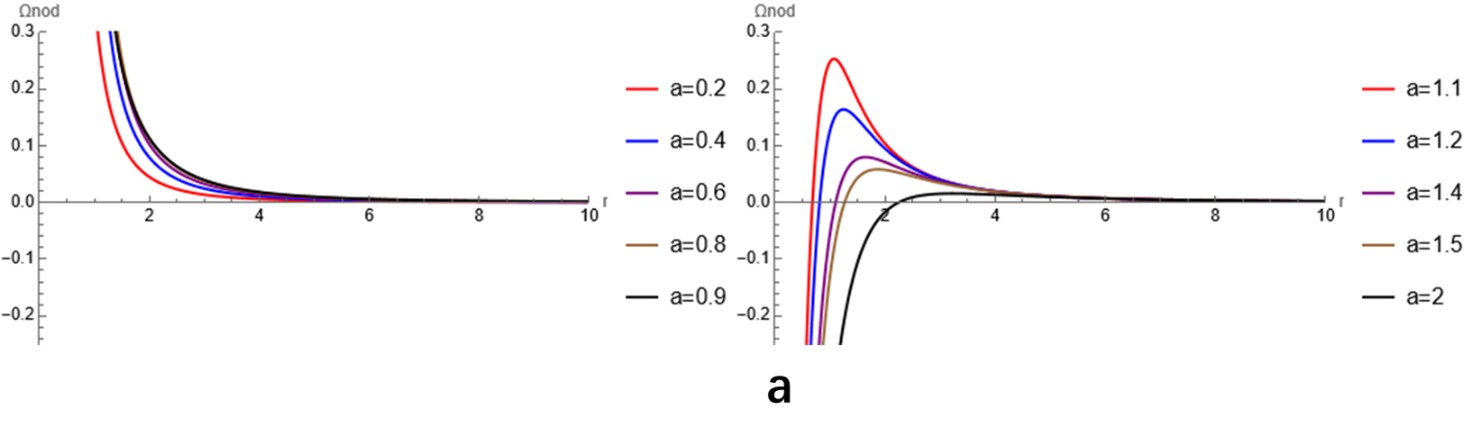}
	\end{minipage}
	\hfill
	\begin{minipage}[b]{0.5\textwidth}
		\includegraphics[width=\textwidth, height=5cm, keepaspectratio]{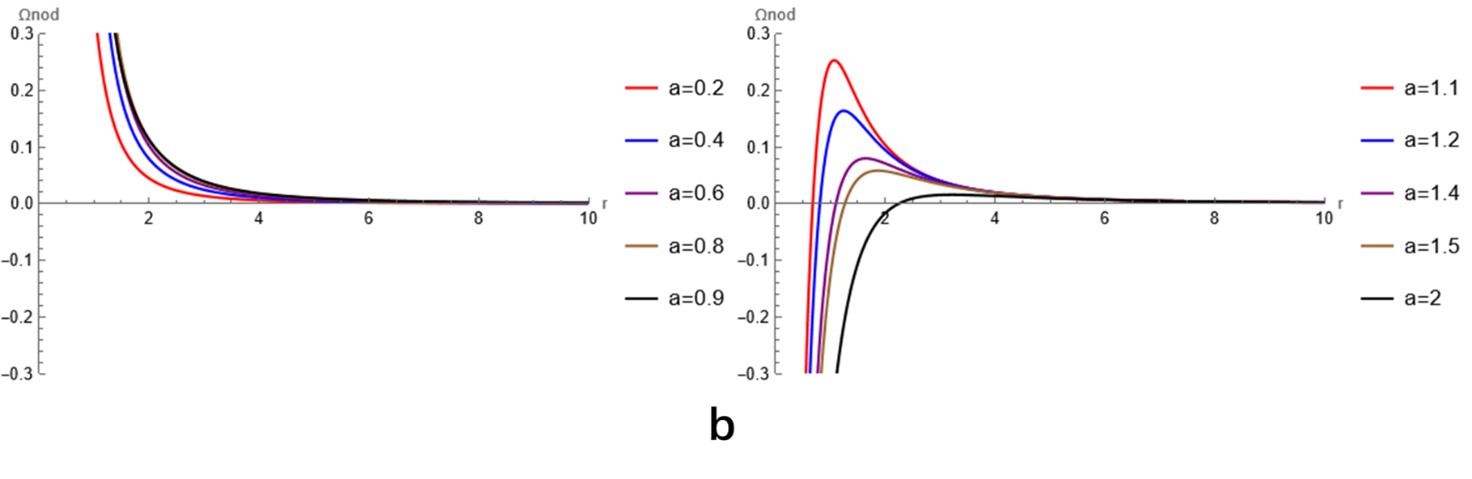}
	\end{minipage}
	\caption{The figure shows the variation of $\Omega_{\text{nod}}$ with $r$ for the cold dark matter (CDM) case (a) and the scalar field dark matter (SFDM) case (b).}\label{fig:15}
\end{figure}
In Fig.~\ref{fig:15}, for both the CDM and SFDM cases, when the singularity is a black hole (BH), $\Omega_{\text{nod}}$ decreases monotonically with increasing $r$. When the singularity is a naked singularity (NS), the $\Omega_{\text{nod}}$ curve exhibits a peak, and the peak value decreases as the spin parameter $a$ increases, while the location of the peak ($r$) shifts to larger $r$ as $a$ increases. Based on the results calculated from Table~\ref{tab:sfdm} and Table~\ref{tab:cdm}, it is evident that at $r = r_{\text{ISCO}}$, $(\partial r_{\text{ISCO}} / \partial r) > 0$.
\subsection{$\Omega_{per}$}
\begin{figure}[htbp]
	\centering
	\begin{minipage}[b]{0.5\textwidth}
		\includegraphics[width=\textwidth, height=5cm, keepaspectratio]{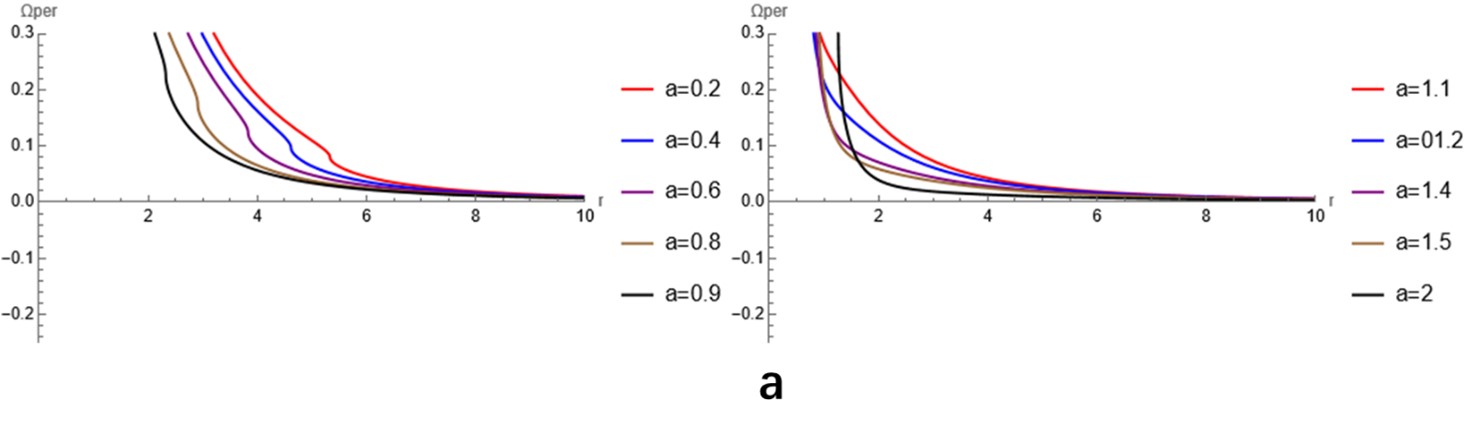}
	\end{minipage}
	\hfill
	\begin{minipage}[b]{0.5\textwidth}
		\includegraphics[width=\textwidth, height=5cm, keepaspectratio]{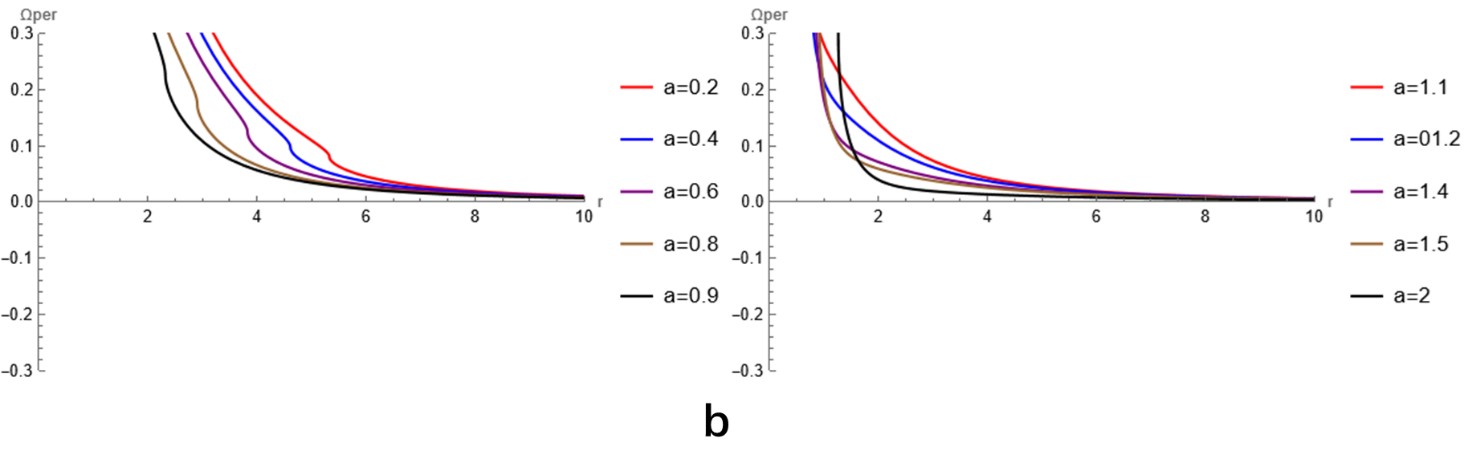}
	\end{minipage}
	\caption{As shown in the figure, the variation of $\Omega_{\text{per}}$ with $r$ is plotted for the cold dark matter (CDM) case (a) and the scalar field dark matter (SFDM) case (b), where the value of $\Omega_{\text{per}}$ decreases monotonically with increasing $r$.}\label{fig:16}
\end{figure}
\section*{Observational Aspects}

Before exploring the physical mechanisms of QPOs in dark matter-modified spacetimes, we first review the fundamental observational properties and theoretical framework of QPOs. Rapid X-ray variability is widely observed in the X-ray emissions from accreting compact objects (such as black holes and neutron stars) e.g.\cite{Remillard2006,vanderKlis2006}, among which quasi-periodic oscillations (QPOs) are a key probe for studying the dynamics in strong gravitational fields due to their prominent narrow-peak power spectrum characteristics\cite{McClintock2001}. Since the Rossi X-ray Timing Explorer (RXTE) systematically measured QPO signals in black hole X-ray binaries (BHXBs) for the first time e.g.\cite{Remillard1999,van_der_Klis2000,Bradt1993}, their frequency coverage (from millihertz to hundreds of hertz) and classification system have been widely established e.g.\cite{Remillard2006}. Current mainstream theories suggest that QPOs are closely related to orbital resonance mechanisms in the inner regions of accretion flows, In particular, the Lense-Thirring (LT) precession induced by frame-dragging could potentially contribute to the observed frequency ratio characteristics of high-frequency QPOs (HFQPOs, 50–450 Hz), as suggested by some theoretical models and numerical simulations e.g.\cite{Ingram2009,Fragile2016}.

Based on their frequency distribution and variability properties, QPOs can be divided into two categories:
\begin{itemize}
	\item \text{High-Frequency QPOs (HFQPO):} Frequency range 50--450 Hz.
	\item \text{Low-Frequency QPOs (LFQPO):} Divided into A, B, and C types with frequency ranges:
	\begin{itemize}
		\item Type A: 6.5--8 Hz,
		\item Type B: 0.8--6.4 Hz,
		\item Type C: 0.01--30 Hz.
	\end{itemize}
\end{itemize}

According to the orbital dynamics model in general relativity, the motion of particles in an accretion disk can be characterized by three fundamental frequencies:
\begin{align}
	\nu_\theta &= f_\theta (r, M, a), \\
	\nu_r &= f_r (r, M, a), \\
	\nu_{\text{nod}} &= f_{\text{nod}} (r, M, a).
\end{align}

Combining the two dark matter models (cold dark matter model CDM and scalar field dark matter model SFDM), numerical calculations of these three frequencies at the innermost stable circular orbit (ISCO) radius, $r_{\text{ISCO}}$, were performed. For a compact object with $M = 10 M_\odot$, the natural unit values are converted to the SI system using the formula:
\begin{align}
	\nu_\alpha = \frac{\Omega_\alpha}{2\pi} \cdot 2.03 \cdot 10^3 \quad (\alpha = \theta, r, \text{nod}, \phi).
\end{align}
\begin{figure}[htbp]
	\centering
	\begin{minipage}[b]{0.5\textwidth}
		\includegraphics[width=\textwidth, height=5cm, keepaspectratio]{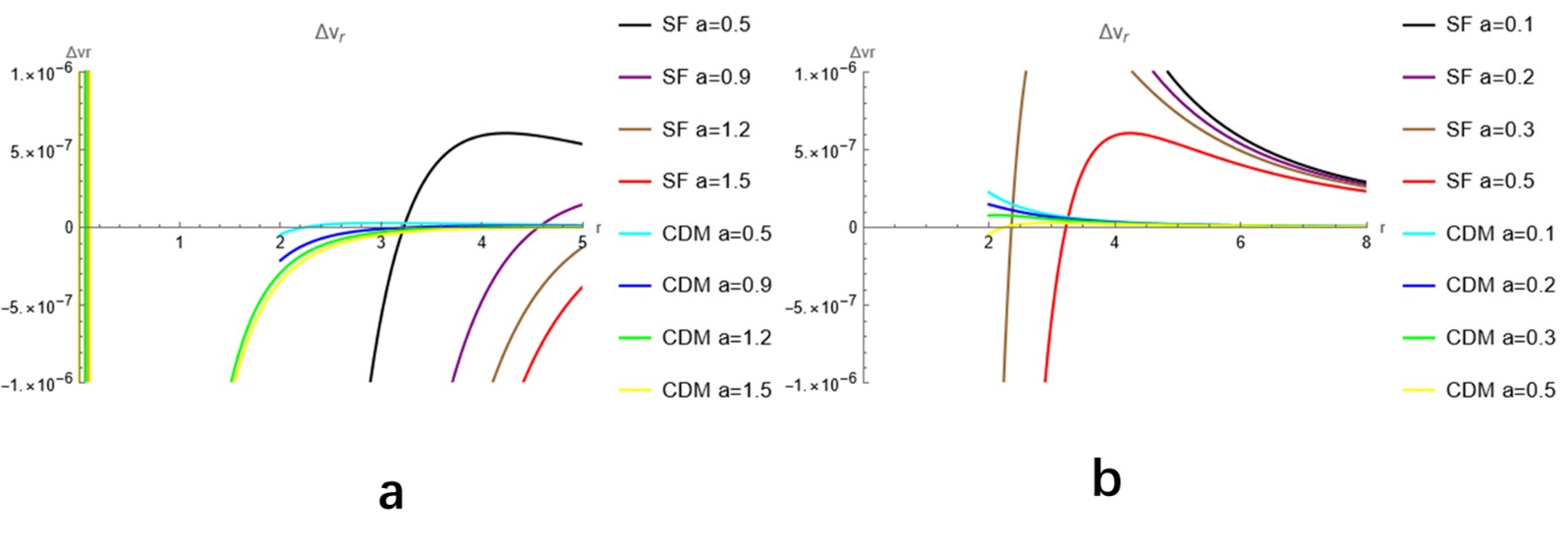}
	\end{minipage}
	\hfill
	\begin{minipage}[b]{0.5\textwidth}
		\includegraphics[width=\textwidth, height=5cm, keepaspectratio]{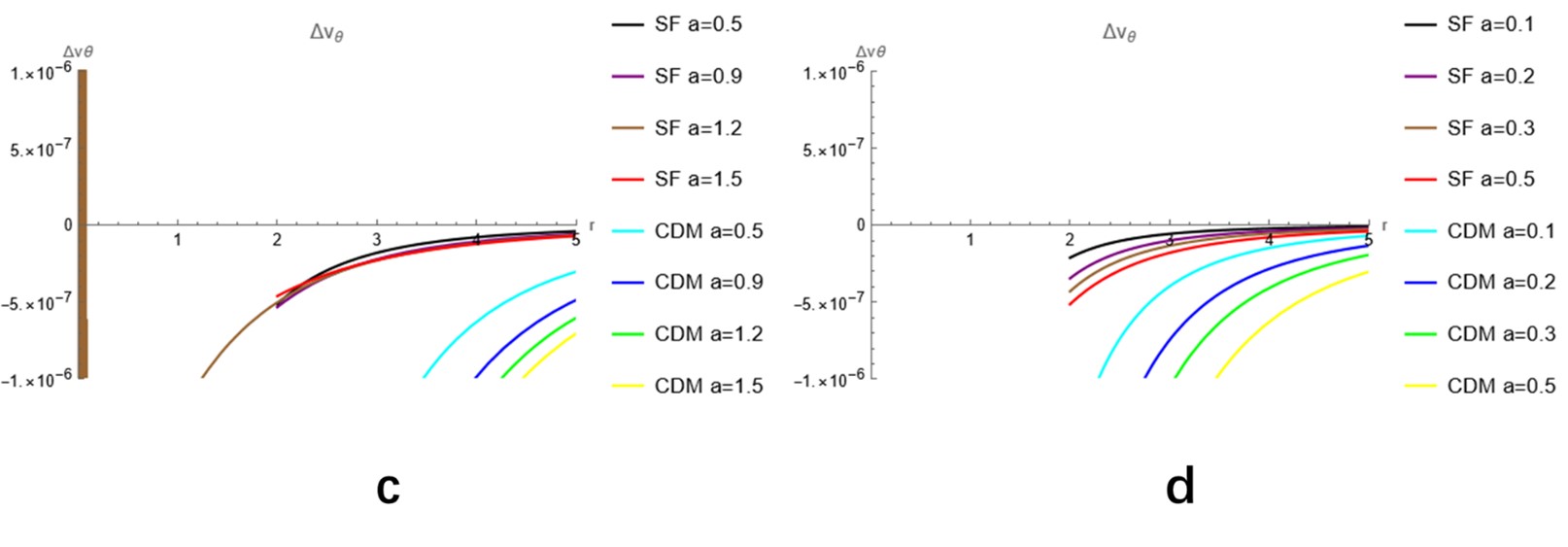}
	\end{minipage}
	\begin{minipage}[b]{0.5\textwidth}
		\includegraphics[width=\textwidth, height=5cm, keepaspectratio]{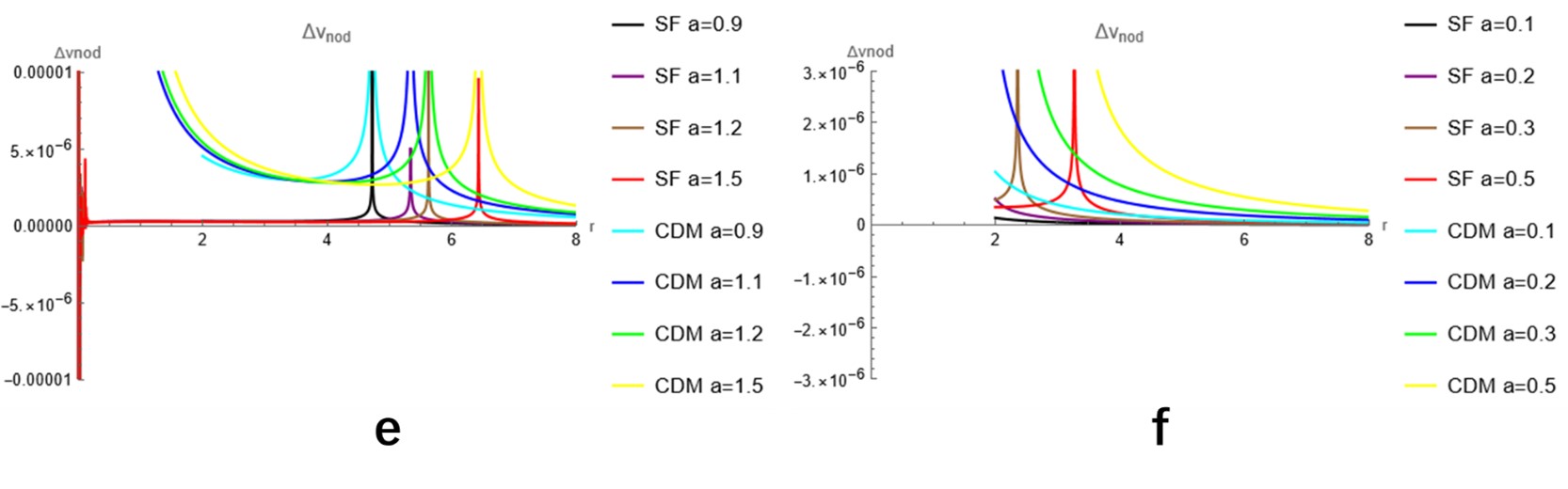}
	\end{minipage}
	\caption{We plotted the differences in the three characteristic frequencies \(\nu_r\), \(\nu_\theta\), and \(\nu_{\text{nod}}\) between CDM and SFDM and the Kerr spacetime at the equatorial plane (\(\theta = \pi/2\)), denoted as \(\Delta\nu_r\) (Figures a, b), \(\Delta\nu_\theta\) (Figures c, d), and \(\Delta\nu_{\text{nod}}\) (Figures e, f). For the black hole (BH) case, the parameters \(a\) were chosen as 0.1, 0.2, 0.3, 0.4, 0.5, and 0.9, while for the neutron star (NS) case, the parameters \(a\) were chosen as 1.1, 1.2, and 1.5.
	}\label{fig:17}
\end{figure}
From Fig.~17, we observe that the influence of dark matter (SFDM and CDM) on the characteristic frequencies is on the order of \(10^{-6}\). The effect of SFDM on \(\nu_r\) and \(\nu_\theta\) is significantly stronger than that of CDM, while its effect on \(\Delta\nu_{\text{nod}}\) is weaker than that of CDM.
\section{Conclusion}
\textbf{1. Theoretical Feature Comparison Between Black Holes and Naked Singularities} \\
\textbf{1.1 Differences in Precession Frequencies Inside and Outside the Ergosphere} \\
\textit{Black Hole (BH):} \\
\begin{itemize}
	\item The Lense-Thirring (LT) precession frequency tends to vanish near the horizon and exhibits a single peak at the static limit surface (\(g_{tt} = 0\)).
	\item The precession frequency outside the ergosphere shows a disappearing peak when the parameter \(q = 0.5\).
\end{itemize}
\textit{Naked Singularity (NS):} \\
\begin{itemize}
	\item The precession frequency near \(r \to 0\) (close to the singularity) remains finite except at the equatorial plane (\(\theta = \pi/2\)), where it diverges.
	\item On the equatorial plane, the precession frequency exhibits a single peak outside the singularity.
	\item For \(\theta \neq \pi/2\), the precession frequency displays a double-peak structure, with the separation decreasing as the rotation parameter \(a\) increases or \(\theta\) approaches the equatorial plane, eventually reducing to a single peak near \(r \approx 1\).
	\item When \(\Omega \neq 0\) and \(q = 0.5\), the precession frequency peak disappears.
\end{itemize}

\textbf{1.2 Comparison of Characteristic Frequencies at the Innermost Stable Circular Orbit (ISCO)} \\
\begin{tabular}{|l|p{2.8cm}|p{2cm}|}
	\hline
	\textbf{Frequency Type} & \textbf{Black Hole (BH) Behavior} & \textbf{Naked Singularity (NS) Behavior} \\
	\hline
	Radial Frequency (\(\nu_r\)) & Negative value (retrograde oscillation), within the high-frequency quasi-periodic oscillation (HFQPO, 100--450 Hz) range & Exceeds the HFQPO detection range \\
	\hline
	Vertical Frequency (\(\nu_\theta\)) & Within the HFQPO range; for an extreme BH (\(a = 1\)), it reduces to the low-frequency QPO (LFQPO, <30 Hz) range & Exceeds the HFQPO detection range \\
	\hline
	Nodal Frequency (\(\nu_{\text{nod}}\)) & For \(a \leq 0.4\), within the LFQPO range; for \(a > 0.4\), enters the HFQPO range; extreme BH shows ultra-high-frequency signals & Decreases from ultra-high frequency to the HFQPO range with increasing \(a\) \\
	\hline
\end{tabular}

\textbf{2. Modulation Effects of Dark Matter Models on Characteristic Frequencies} \\
\textbf{2.1 Magnitude and Differences of Influence} \\
\begin{itemize}
	\item \textit{LT Precession Frequency:} The correction magnitude for both CDM and SFDM is on the order of \(10^{-6}\), but CDM modulates the LT precession more strongly than SFDM.
	\item \textit{ISCO Characteristic Frequencies:} \\
	\begin{itemize}
		\item \textit{CDM:} Stronger correction to the nodal frequency (\(\nu_{\text{nod}}\)).
		\item \textit{SFDM:} More significant correction to the radial (\(\nu_r\)) and vertical (\(\nu_\theta\)) frequencies.
	\end{itemize}
\end{itemize}
In general, the precession frequencies near black holes and naked singularities under a dark matter halo exhibit significant differences, particularly at the innermost stable circular orbit. Meanwhile, the modulation effects of dark matter models (CDM and SFDM) on the precession frequencies are concentrated at the \(10^{-6}\) order of magnitude, posing certain challenges to current observational techniques.
\section{acknowledgements}
We acknowledge the anonymous referee for a constructive report that has significantly improved this paper.This work was supported by Guizhou Provincial Basic Research Program(Natural Science)(Grant No.QianKeHeJiChu-[2024]Young166), the Special Natural Science Fund of Guizhou University (Grant No.X2022133), the National Natural Science Foundation of China (Grant No.12365008) and the Guizhou Provincial Basic Research Program (Natural Science) (Grant No.QianKeHeJiChu-ZK[2024]YiBan027 and QianKeHeJiChu-ZK[2025]General Program680) .
	
\bibliographystyle{unsrt} 
\bibliography{reference.bib}
\clearpage

\appendix 
\section{precession}
\label{app:precession}
The complete expression for the precession frequency is:
\begin{align*}
	\Omega_p &= \frac{1}{2 \Sigma^2 \sin\theta \Big(1 + 2\Omega \frac{a t \sin^2\theta}{\Sigma^2 - t} + \Omega^2 \frac{\sin^2\theta \big( (r^2 + a^2)^2 - a^2 \Delta \sin^2\theta \big)}{\Sigma^2 - t} \Big)} \notag\\
	&\quad \times ((A_1 + A_2 + A_3)\hat{r} +( B_1 + B_2 + B_3)\hat{\theta}).\label{app:p}
\end{align*}
The expression for the Lense-Thirring(LT) precession frequency is:
\begin{align*}
	\Omega_p &= \frac{1}{2 \Sigma^2 \sin\theta}( A_1 \hat{r}+B_1\hat{\theta}).
\end{align*}
where
\begin{align*}
	A_1 &= \Sigma \left( g_{t\phi,r} - \frac{g_{t\phi}}{g_{tt}} g_{tt,r} \right) \partial_\theta \\
	&= -\frac{a \sin^2\theta}{\Sigma^3} \left( t_r \Sigma^2 - 2r t \right) \left( 1 - \frac{t}{\Sigma^2 - t} \right) \partial_\theta,
\end{align*}
\begin{align*}
	A_2 &= \Omega \Sigma \left( g_{\phi\phi,r} - \frac{g_{\phi\phi}}{g_{tt}} g_{tt,r} \right) \partial_\theta \\
	&= \Omega \Sigma \left[ \frac{\sin^2\theta}{\Sigma^2} \left( 4r (r^2 + a^2) - a^2 \Delta_r \sin^2\theta \right) \right. \\
	&\quad \left. - \frac{2r \sin^2\theta}{\Sigma^4} \left( (r^2 + a^2)^2 - a^2 \Delta \sin^2\theta \right) \right. \\
	&\quad \left. + \frac{\sin^2\theta \left( (r^2 + a^2)^2 - a^2 \Delta \sin^2\theta \right) (t_r \Sigma^2 - 2r t)}{\Sigma^4 \left(1 - \frac{t}{\Sigma^2}\right)} \right] \partial_\theta, 
\end{align*}
\begin{align*}
	A_3 &= \Omega^2 \Sigma \left( \frac{g_{t\phi}}{g_{tt}} g_{\phi\phi,r} - \frac{g_{\phi\phi}}{g_{tt}} g_{t\phi,r} \right) \partial_\theta \\
	&= \Omega^2 \Sigma \left[ \frac{a t \sin^2\theta}{\Sigma^2 \left(1 - \frac{t}{\Sigma^2}\right)} \left( \frac{\sin^2\theta}{\Sigma^2} \left( 4r (r^2 + a^2) - a^2 \Delta_r \sin^2\theta \right) \right. \right. \\
	&\quad \left. - \frac{2r \sin^2\theta}{\Sigma^4} \left( (r^2 + a^2)^2 - a^2 \Delta \sin^2\theta \right) \right) \\
	&\quad \left. + \frac{\sin^2\theta \left( (r^2 + a^2)^2 - a^2 \Delta \sin^2\theta \right) a \sin^2\theta (t_r \Sigma^2 - 2r t)}{\Sigma^4 \left(1 - \frac{t}{\Sigma^2}\right)} \right] \partial_\theta, 
\end{align*}
\begin{align*}
	B_1 &= -\frac{\Sigma}{\sqrt{\Delta}} \left( g_{t\phi,\theta} - \frac{g_{t\phi}}{g_{tt}} g_{tt,\theta} \right) \partial_r \\
	&= -\frac{\Sigma}{\sqrt{\Delta}} \left[ -\frac{2a t \sin\theta \cos\theta}{\Sigma^2} \left( 1 + \frac{a^2 \sin^2\theta}{\Sigma^2} \right) \right. \\
	&\quad \left. + \frac{a t \sin^2\theta \cdot 2a^2 t \sin\theta \cos\theta}{\Sigma^6 \left(1 - \frac{t}{\Sigma^2}\right)} \right] \partial_r, 
\end{align*}
\begin{align*}
	B_2 &= -\Omega \frac{\Sigma}{\sqrt{\Delta}} \left( g_{\phi\phi,\theta} - \frac{g_{\phi\phi}}{g_{tt}} g_{tt,\theta} \right) \partial_r \\
	&= -\Omega \frac{\Sigma}{\sqrt{\Delta}} \left[ \frac{2 \sin\theta \cos\theta}{\Sigma^2} \left( (r^2 + a^2)^2 - a^2 \Delta \sin^2\theta \right) \right. \\
	&\quad - \frac{2a^2 \Delta \sin^3\theta \cos\theta}{\Sigma^2} + \frac{2a^2 \sin^3\theta \cos\theta}{\Sigma^4} \left( (r^2 + a^2)^2 - a^2 \Delta \sin^2\theta \right) \\
	&\quad \left. + \frac{\sin^2\theta \left( (r^2 + a^2)^2 - a^2 \Delta \sin^2\theta \right) 2a^2 t \sin\theta \cos\theta}{\Sigma^6 \left(1 - \frac{t}{\Sigma^2}\right)} \right] \partial_r, 
\end{align*}
\begin{align*}
	B_3 &= -\Omega^2 \frac{\Sigma}{\sqrt{\Delta}} \left( \frac{g_{t\phi}}{g_{tt}} g_{\phi\phi,\theta} - \frac{g_{\phi\phi}}{g_{tt}} g_{t\phi,\theta} \right) \partial_r \\
	&= -\Omega^2 \frac{\Sigma}{\sqrt{\Delta}} \biggl[ \frac{a t \sin^2\theta}{\Sigma^2 \left(1 - \frac{t}{\Sigma^2}\right)} \biggl( \frac{2 \sin\theta \cos\theta}{\Sigma^2} \bigl( (r^2 + a^2)^2 \\
	&\quad - a^2 \Delta \sin^2\theta \bigr) - \frac{2a^2 \Delta \sin^3\theta \cos\theta}{\Sigma^2} \\
	&\quad + \frac{2a^2 \sin^3\theta \cos\theta}{\Sigma^4} \bigl( (r^2 + a^2)^2 - a^2 \Delta \sin^2\theta \bigr) \biggr) \\
	&\quad + \frac{\sin^2\theta \bigl( (r^2 + a^2)^2 - a^2 \Delta \sin^2\theta \bigr)}{\Sigma^6 \left(1 - \frac{t}{\Sigma^2}\right)} 2a t \sin\theta \cos\theta \\
	&\quad \times \left( 1 + \frac{a^2 \sin^2\theta}{\Sigma^2} \right) \biggr] \partial_r
\end{align*}
For CDM case:
\begin{align*}
	t &= r^2 + \frac{2GM}{c^2} r - r^2 \left[1 + \frac{r}{R_s}\right]^{-\frac{8\pi G \rho_c R_s^3}{c^2 r}}, \\
	\Delta &= r^2 \left[1 + \frac{r}{R_s}\right]^{-\frac{8\pi G \rho_c R_s^3}{c^2 r}} - \frac{2GM}{c^2} r + a^2, 
\end{align*}
\begin{align*}
	t_r &= 2r + \frac{2GM}{c^2} - 2r \left[1 + \frac{r}{R_s}\right]^{-\frac{8\pi G \rho_c R_s^3}{c^2 r}} \\
	&\quad + r^2 \frac{8\pi G \rho_c R_s^3}{c^2 r^2} \left[1 + \frac{r}{R_s}\right]^{-\frac{8\pi G \rho_c R_s^3}{c^2 r} - 1} \\
	&\quad \times \left( \frac{1}{R_s} + \frac{8\pi G \rho_c R_s^3}{c^2 r^2} \right), 
\end{align*}
\begin{align*}
	\Delta_r &= 2r \left[1 + \frac{r}{R_s}\right]^{-\frac{8\pi G \rho_c R_s^3}{c^2 r}} - \frac{2GM}{c^2} \\
	&\quad - r^2 \frac{8\pi G \rho_c R_s^3}{c^2 r^2} \left[1 + \frac{r}{R_s}\right]^{-\frac{8\pi G \rho_c R_s^3}{c^2 r} - 1} \\
	&\quad \times \left( \frac{1}{R_s} + \frac{8\pi G \rho_c R_s^3}{c^2 r^2} \right)
\end{align*}
For SFDM case:
\begin{align*}
	t &= r^2 + \frac{2GM}{c^2} r - r^2 \exp\left[ -\frac{8G \rho_c R_s^2}{\pi} \frac{\sin\left( \frac{\pi r}{R_s} \right)}{ \frac{\pi r}{R_s} } \right], \\
	\Delta &= r^2 \exp\left[ -\frac{8G \rho_c R_s^2}{\pi} \frac{\sin\left( \frac{\pi r}{R_s} \right)}{ \frac{\pi r}{R_s} } \right] - \frac{2GM}{c^2} r + a^2, 
\end{align*}
\begin{align*}
	t_r &= 2r + \frac{2GM}{c^2} - 2r \exp\left[ -\frac{8G \rho_c R_s^2}{\pi} \frac{\sin\left( \frac{\pi r}{R_s} \right)}{ \frac{\pi r}{R_s} } \right] \\
	&\quad + r^2 \exp\left[ -\frac{8G \rho_c R_s^2}{\pi} \frac{\sin\left( \frac{\pi r}{R_s} \right)}{ \frac{\pi r}{R_s} } \right] \\
	&\quad \times \frac{8G \rho_c R_s^2}{\pi} \left( -\frac{\pi}{R_s} \frac{\cos\left( \frac{\pi r}{R_s} \right)}{ \frac{\pi r}{R_s} } + \frac{\sin\left( \frac{\pi r}{R_s} \right)}{ \frac{\pi r^2}{R_s} } \right), 
\end{align*}
\begin{align*}
	\Delta_r &= 2r \exp\left[ -\frac{8G \rho_c R_s^2}{\pi} \frac{\sin\left( \frac{\pi r}{R_s} \right)}{ \frac{\pi r}{R_s} } \right] - \frac{2GM}{c^2} \\
	&\quad - r^2 \exp\left[ -\frac{8G \rho_c R_s^2}{\pi} \frac{\sin\left( \frac{\pi r}{R_s} \right)}{ \frac{\pi r}{R_s} } \right] \\
	&\quad \times \frac{8G \rho_c R_s^2}{\pi} \left( -\frac{\pi}{R_s} \frac{\cos\left( \frac{\pi r}{R_s} \right)}{ \frac{\pi r}{R_s} } + \frac{\sin\left( \frac{\pi r}{R_s} \right)}{ \frac{\pi r^2}{R_s} } \right)
\end{align*}

\end{document}